\documentclass[10pt,twocolumn,letterpaper]{article}
\pdfoutput=1
\usepackage{iccv}              
\definecolor{iccvblue}{rgb}{0.21,0.49,0.74}
\usepackage[pagebackref,breaklinks,colorlinks,allcolors=iccvblue]{hyperref}

\usepackage{booktabs, multirow, tabularx}
\usepackage{siunitx} 
\usepackage{makecell}
\usepackage{microtype}
\usepackage{graphicx}
\usepackage{mathrsfs}
\usepackage{algorithm}
\usepackage{algorithmic}
\usepackage{xspace}
\usepackage{arydshln}    
\usepackage{array}       
\usepackage{bigstrut}

\newcolumntype{C}{>{\centering\arraybackslash}X}

\newcommand{\name}{\textsc{Burn}\xspace}
\newcommand{\Tref}[1]{Table~\ref{#1}}

\newcommand{\Fref}[1]{Fig.~\ref{#1}}

\newcommand{\Aref}[1]{Alg.~\ref{#1}}

\def\@onedot{\ifx\@let@token.\else.\null\fi\xspace}

\def\ie{\emph{i.e}\onedot}

\def\paperID{15662} 
\def\confName{ICCV}
\def\confYear{2025}

\title{\name: Backdoor Unlearning via Adversarial Boundary Analysis}
\author{
Yanghao Su$^1$, Jie Zhang$^2$\thanks{The corresponding author: Jie Zhang (zhang\_jie@cfar.a-star.edu.sg)}, Yiming Li$^3$, Tianwei Zhang$^3$\\
Qing Guo$^2$, Weiming Zhang$^1$, Nenghai Yu$^1$, Nils Lukas$^4$, Wenbo Zhou$^1$ \\[0.1ex]
$^1$University of Science and Technology of China   \\
$^2$CFAR and IHPC, A*STAR   \\
$^3$Nanyang Technological University   \\
$^4$Mohamed bin Zayed University of Artificial Intelligence   \\
}

\begin{document}
\maketitle

\begin{abstract}
Backdoor unlearning aims to remove backdoor-related information while preserving the model's original functionality. However, existing unlearning methods mainly focus on recovering trigger patterns but fail to restore the correct semantic labels of poison samples. This limitation prevents them from fully eliminating the false correlation between the trigger pattern and the target label.
To address this, we leverage boundary adversarial attack techniques, revealing two key observations. 
First, poison samples exhibit significantly greater distances from decision boundaries compared to clean samples, indicating they require larger adversarial perturbations to change their predictions.
Second, while adversarial predicted labels for clean samples are uniformly distributed, those for poison samples tend to revert to their original correct labels.
Moreover, the features of poison samples restore to closely resemble those of corresponding clean samples after adding adversarial perturbations.

Building upon these insights, we propose \textbf{B}ackdoor \textbf{U}nlearning via adversa\textbf{R}ial bou\textbf{N}dary analysis (\name), a novel defense framework that integrates false correlation decoupling, progressive data refinement, and  model purification.
In the first phase, \name employs adversarial boundary analysis to detect poisoned samples based on their abnormal adversarial boundary distances, then restores their correct semantic labels for fine-tuning. In the second phase, it employs a feedback mechanism that tracks prediction discrepancies between the original backdoored model and progressively sanitized models, guiding both dataset refinement and model purification.
Extensive evaluations across multiple datasets, architectures, and seven diverse backdoor attack types confirm that \name effectively removes backdoor threats while maintaining the model’s original performance.
\end{abstract}




\vspace{-1em}
\section{Introduction}
\label{submission}
Deep Neural Networks (DNNs) are vulnerable to backdoor attacks \cite{gu2017badnets, li2021invisible, zeng2021rethinking, wang2022bppattack, liu2018trojaning, doan2021lira, bagdasaryan2021blind}, where an adversary injects a backdoor trigger (e.g., a pixel patch) into a small subset of training samples and mislabels them as a \emph{target} class.
These manipulated samples, known as \textit{poison samples}, collectively form a poisoned dataset. As a result, the model learns a backdoor correlation between the trigger and the target class while still performing normally under standard evaluation metrics. This vulnerability poses a serious threat to security-sensitive applications, including but not limited to face recognition \cite{parkhi2015deep}, biomedical diagnosis \cite{esteva2017dermatologist}, and autonomous driving \cite{redmon2016you}.

\begin{figure}[t]
	\centering
 	\includegraphics[width=.9\linewidth] {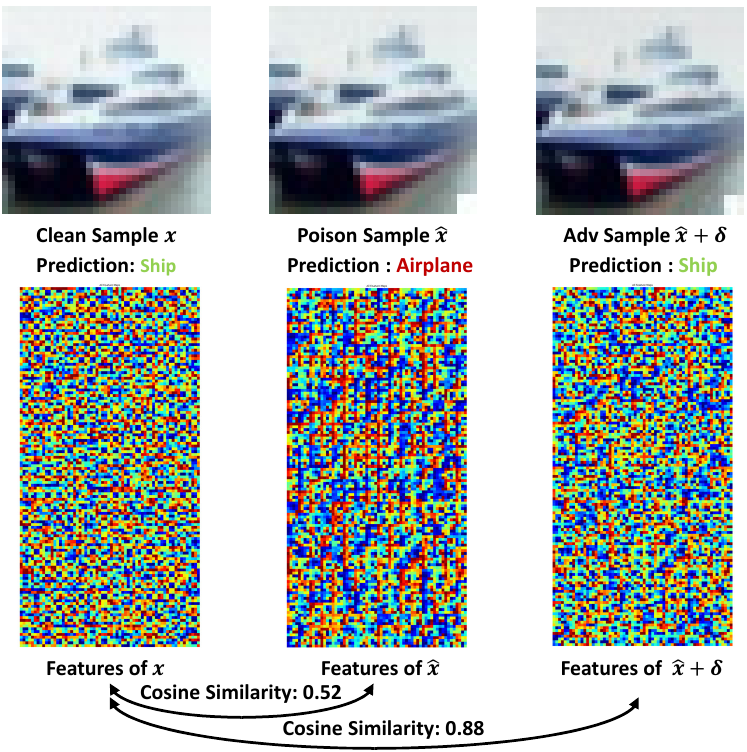}\label{fig:feature} 
	\caption{Predicted labels and features of clean sample $x$, poison sample $\hat{x}$, and adversarially perturbed poison sample $\hat{x}+\delta$. Boundary adversarial attacks effectively restore correct predictions and align poison sample features with clean ones, as shown by increased cosine similarity.}
	\label{fig:feature} 
\end{figure}

Backdoor unlearning aims to reverse this process by removing the influence of backdoor triggers, thereby removing malicious behaviors from a compromised model while preserving its original performance.
However, existing unlearning methods primarily focus on recovering trigger patterns, which becomes increasingly challenging as attackers adopt stealthy (e.g., steganography-based \cite{li2021invisible}) and dynamic (e.g., input-dependent \cite{nguyen2020input}) triggers. These approaches struggle to fully eliminate the learned correlation between the trigger and target label, limiting their effectiveness \cite{wang2019neural,li2021neural}.
To address these limitations, we propose Backdoor Unlearning via Adversarial Boundary Analysis (\name), a unified defense framework that integrates false correlation decoupling and progressive model purification. Instead of relying on trigger pattern recovery, \name directly restores the correct labels of poisoned examples, effectively breaking the backdoor association. This novel approach makes \name more robust against advanced backdoor attacks and enhances its adaptability to complex poisoning strategies.


Our approach is motivated by insights gained from observations based on boundary adversarial attacks \cite{moosavi2016deepfool,abroshan2025superdeepfool}. These attacks generate minimal perturbations that push input samples across classifier decision boundaries by exploiting the geometric properties of deep neural networks, effectively crafting subtle yet impactful adversarial examples.
By applying boundary adversarial attacks, we uncover two critical observations: 1) \textbf{Poison samples exhibit significantly greater distances from decision boundaries} compared to clean samples, as demonstrated in \Tref{tab:boundary}. This is evident from the much larger perturbation magnitudes (measured by \(L_2\) norms) required to induce misclassification.
2) \textbf{Adversarial predicted labels behave differently for poison and clean samples}. As shown in \Fref{fig:labels}, while adversarial labels for clean samples are uniformly distributed, those for poisoned samples overwhelmingly revert to their original, correct labels.
To further analyze these effects, we visualize and compare the feature representations of (i) a clean sample x, (ii) its corresponding poison sample $\hat{x}$, and (iii) the boundary adversarial sample $\hat{x+}\delta$ generated from $\hat{x}$. As illustrated in \Fref{fig:feature}, after applying adversarial perturbations $\delta$, the features of the poison sample $\hat{x}$ closely resemble those of the corresponding clean sample x. These findings indicate that boundary adversarial attacks not only expose the abnormal decision boundary distances of poison samples but also demonstrate the ability to ``strip away” trigger-dominated features, restoring the original semantic structure of the poison samples. This effect is evident both in the predicted label distributions and in feature space.


Our method leverages adversarial boundary analysis and a dual-model prediction discrepancy feedback mechanism to iteratively refine poison data and purify the model. This process effectively decouples false correlations introduced by backdoor triggers while preserving the model’s performance on clean tasks.
In the first phase, we aim to identify and correct poison samples. Specifically, we apply boundary adversarial attack analysis to detect suspicious poison samples based on their abnormal boundary distances. These samples are then re-labeled with their correct semantic labels.
In the second phase, we conduct progressive data refinement and model purification. 
Using the re-labeled poison samples, we fine-tune the model while keeping the original infected model fixed. By continuously applying boundary adversarial attack analysis to the frozen infected model and measuring the discrepancies in predictions between the infected and progressively refined models, we iteratively enhance the dataset quality and purify the model.




In summary, our contributions are as follows:

\begin{itemize}
\item We uncover two key insights through adversarial boundary analysis:
1) Poison samples exhibit significantly greater distances from decision boundaries than clean samples, as indicated by the larger perturbation magnitudes ($L\_2$ norms) required for misclassification. 2) Adversarial labels for poison samples tend to revert to their original correct labels, unlike clean samples, whose adversarial labels are uniformly distributed.
Additionally, applying adversarial perturbations restores poison sample features to closely resemble those of their clean counterparts.

\item  We introduce \name, a novel defense framework that decouples false backdoor correlations and integrates progressive data refinement with model purification. Unlike prior approaches, \name operates without relying on assumptions about specific trigger patterns and without requiring a clean auxiliary dataset.

\item We validate \name across multiple datasets, architectures, and seven different backdoor attacks with diverse trigger types, demonstrating its effectiveness and robustness.



\end{itemize}

\section{Related Work \& Preliminary}
\subsection{Backdoor Attacks}

The landscape of backdoor attacks on deep neural networks (DNNs) has evolved significantly, with increasing stealth and effectiveness in trigger design. The seminal work BadNets \cite{gu2017badnets} first demonstrated DNN vulnerabilities using visible square triggers. Subsequent research introduced various visible trigger techniques, such as image blending in the Blended attack \cite{chen2017targeted} and frequency-domain triggers in the Low-Frequency (LF) attack \cite{zeng2021rethinking}.
To enhance stealth, researchers developed imperceptible trigger patterns, including image quantization and dithering in BPP \cite{wang2022bppattack} and steganography-based triggers in SSBA \cite{li2021invisible}, and triggers that use image structures to inject information \cite{zhang2022poison}. More recently, clean-label backdoor attacks have emerged, where the poison samples retain their correct ground-truth labels, avoiding obvious labeling anomalies and increasing attack stealth.
Additionally, backdoor triggers have evolved from static to dynamic patterns \cite{nguyen2020input,li2021invisible,doan2021lira,schneider2024universal,zhang2022poison}. Dynamic triggers adapt to individual input samples, making them more effective and harder to detect compared to fixed-pattern triggers.


Beyond data poisoning-based attacks, another category of backdoor attacks leverages model modification techniques. TrojanNN \cite{liu2018trojaning} optimizes triggers to activate crucial neurons at maximal values, while LIRA \cite{doan2021lira} formulates a non-convex, constrained optimization problem to learn invisible triggers through a two-stage stochastic optimization process.


\subsection{Backdoor Defense}
Backdoor defenses in deep neural networks (DNNs) are broadly classified into two categories: detection-based and mitigation-based methods.
Detection-based approaches aim to identify backdoors in models or filter out suspicious samples from training data for retraining. While effective in detecting compromised models, these methods do not remove the backdoor itself.
Mitigation-based approaches focus on purifying compromised models by reducing the influence of backdoor triggers while preserving clean data performance.
One common technique is fine-tuning with a clean auxiliary dataset.
Neural pruning methods, such as Fine-Pruning \cite{liu2018fine}, remove neurons associated with backdoor behavior.
Trigger synthesis techniques, such as Neural Cleanse (NC) \cite{wang2019neural} and Artificial Brain Stimulation (ABS) \cite{liu2019abs}, reconstruct and leverage backdoor triggers for purification.
Knowledge distillation methods like Neural Attention Distillation (NAD) \cite{li2021neural} erase backdoors by transferring knowledge from a compromised model to a new clean model.
Adversarial perturbation techniques, such as Adversarial Neuron Pruning (ANP) \cite{wu2021adversarial}, prune backdoor-related neurons by perturbing model weights.
Despite their effectiveness, many of these methods assume fixed triggers and struggle against dynamic or content-aware backdoor attacks, where triggers adapt based on input content (e.g., Dynamic Backdoor Attacks).


Recent research has introduced more robust and adaptive defense mechanisms:
Causality-inspired Backdoor Defense (CBD) \cite{xiang2024cbd} learns deconfounded representations to mitigate backdoor effects without requiring a clean dataset.
Decoupling-Based Defense (DBD) \cite{huang2022backdoor} separates different training components to reduce backdoor impact.
Implicit Backdoor Adversarial Unlearning (I-BAU) \cite{zeng2021adversarial} detects potential backdoor triggers using universal adversarial attacks and unlearns them to purify the model.
Shared Adversarial Unlearning (SAU) \cite{wei2023shared} establishes a link between backdoor risk and adversarial risk, proposing a bi-level optimization framework that leverages adversarial training for backdoor mitigation.
Progressive Backdoor Erasing (PBE) \cite{mu2023progressive} exploits similarities between adversarial examples and backdoor-triggered images to cleanse models.
FT-SAM \cite{zhu2023enhancing} enhances fine-tuning for backdoor removal by incorporating sharpness-aware minimization, improving model robustness.
    

\subsection{Boundary Adversarial Attacks}
Drawing on previous studies\cite{moosavi2016deepfool,abroshan2025superdeepfool}, in this section, we formalize the boundary adversarial attack.

For a binary linear classifier \( f(x) = w^T x + b \), the decision boundary is defined by the hyperplane:  
\[
\mathscr{F} = \left\{ x \mid w^T x + b = 0 \right\},
\]  
with normal vector \( w \). The minimal perturbation required to misclassify a given sample \( x_0 \) is given by:  
\[
d(x_0) = \frac{|w^T x_0 + b|}{\|w\|_2}.
\]

For a \( K \)-class linear classifier, the boundary between class \( k \) and the predicted class \( \hat{k} \) is defined by:  
\[
(w_k - w_{\hat{k}})^T x + (b_k - b_{\hat{k}}) = 0,
\]  
with normal vector \( w_k - w_{\hat{k}} \). The distance from \( x_0 \) to this boundary is:  
\[
d_k(x_0) = \frac{|f_k(x_0) - f_{\hat{k}}(x_0)|}{\|w_k - w_{\hat{k}}\|_2}.
\]  
The smallest such distance determines the optimal adversarial direction:  
\[
\delta^*(x_0) = \frac{f_{\hat{k}}(x_0) - f_{k^*}(x_0)}{\|w_{\hat{k}} - w_{k^*}\|_2^2} (w_{\hat{k}} - w_{k^*}),
\]  
where \( k^* = \arg \min_{k \neq \hat{k}} d_k(x_0) \), corresponding to the closest decision boundary.

For nonlinear deep models, weight vectors are no longer explicitly defined, so gradients are used to approximate local decision boundaries. 
For a sample \( x_0 \), the first-order Taylor expansion gives:  
\[
f_k(x_0 + \delta) - f_{\hat{k}}(x_0 + \delta) \approx (\nabla f_k - \nabla f_{\hat{k}})^T \delta,
\]  
where \( \nabla f_k \) represents the gradient of the \( k \)-th class score. This motivates using gradient differences \( \nabla f_k - \nabla f_{\hat{k}} \) as an approximation of the boundary normal.

For a \( K \)-class deep neural network \( f(\cdot) \), the algorithm starts with the original sample \( x_0 \). At each iteration, it computes the gradients \( \nabla f_k(x) \) of all class scores and identifies the closest decision boundary via:  
\[
k^* = \arg \min_{k \neq \hat{k}} \frac{|f_k(x) - f_{\hat{k}}(x)|}{\|\nabla f_k(x) - \nabla f_{\hat{k}}(x)\|_2},
\]  
where \( \hat{k} \) is the predicted class of the current sample \( x \). The perturbation is then calculated as:  
\[
\Delta x = \eta \cdot \frac{f_{\hat{k}}(x) - f_{k^*}(x)}{\|\nabla f_{\hat{k}}(x) - \nabla f_{k^*}(x)\|_2^2} \left( \nabla f_{k^*}(x) - \nabla f_{\hat{k}}(x) \right),
\]  
and the sample is updated as \( x \gets x + \Delta x \). The process terminates when the perturbed sample \( x \) is misclassified.


\begin{table*}[t]
  \centering
  \footnotesize
  \setlength\tabcolsep{7pt}
  \vspace{-1em}
  \caption{Average adversarial boundary distance (measured by $L_2$ norms) for 1000 clean samples and 1000 poison samples across CIFAR-10, CIFAR-100, and Tiny ImagNet-200 . Results are reported from models injected with seven different backdoor attack types.  Poison samples exhibit higher adversarial decision boundary distances compared to clean samples.}
\begin{tabular}{c|c|c|c|c|c|c|c|c}
\hline
  & \textbf{Clean} & \textbf{BadNets} & \textbf{Blended} & \textbf{Dynamic} & \textbf{CL} & \textbf{LF} & \textbf{SSBA} & \textbf{TrojanNN} \bigstrut\\
\hline
CIFAR-10 & 0.571  & 1.116  & 1.227  & 1.047  & 2.151  & 1.312  & 0.858  & 2.812  \bigstrut\\
\hline
CIFAR-100 & 0.261  & 0.795  & 0.913  & 0.877  & 3.693  & 0.690  & 0.631  & 2.650  \bigstrut\\
\hline
Tiny ImagNet-200 & 0.472  & 2.417  & 3.391  & 1.907  & 4.809  & 3.131  & 3.922  & 4.588  \bigstrut\\
\hline
\end{tabular}%
  \label{tab:boundary}%
\end{table*}%

\section{Threat Model}
\subsection{Adversary Capabilities and Goals}
We consider a realistic backdoor attack scenario where an adversary has full control over the training data source, allowing arbitrary modifications such as inserting visible or hidden triggers and relabeling poison samples to arbitrary target classes. However, the attacker lacks access to the model architecture, parameters, and training process.

The adversary's objective is to train an infected classifier $f$ with parameters $\theta$ such that:
\[
\begin{array}{r}\theta=\arg \min _\theta \mathbb{E}_{(x, y) \sim {D}} \mathcal{L}(f(x ; \theta), y) \\ +\mathbb{E}_{\left(\hat{x}, y_\text{target}\right) \sim {\hat{D}}} \mathcal{L}(f(\hat{x} ; \theta), y_\text{target}),
\end{array}
\]
where $D=\left\{\left({x}_i, {y}_i\right)\right\}_{i=1}^n$ and $\hat{D}=\left\{\left(\hat{x}_i, y_\text{target}\right)\right\}_{i=1}^m$ denote the benign samples and trigger samples, respectively. $\mathcal{L}$ denotes the loss function, e.g., cross-entropy loss.
In this setting, the infected model functions normally on benign samples but yields a specific target prediction $y_\text{target}$ when presented with trigger samples $\hat{x} = x + t$. 

\subsection{Defender Capabilities and Goals}
The defender retains complete administrative authority over the model architecture and training protocols, yet confronts substantial operational uncertainties: they possess neither auditing capabilities for the upstream data acquisition pipeline nor prior knowledge of critical attack parameters, including poisoning intensity, target class distributions, or trigger pattern specifications. The defender’s dual objectives are to (i) preserve classification fidelity on benign samples, ensuring parity with uncontaminated model performance, and (ii) prevent targeted misclassification of poison samples, thereby nullifying malicious manipulation.
The attacker has unrestricted control over training data to implant backdoors, while the defender must mitigate these threats without degrading model utility, countering unknown attack strategies and preserving clean-task performance.

\begin{figure}[t]
	\centering
        \includegraphics[width=.7\linewidth] {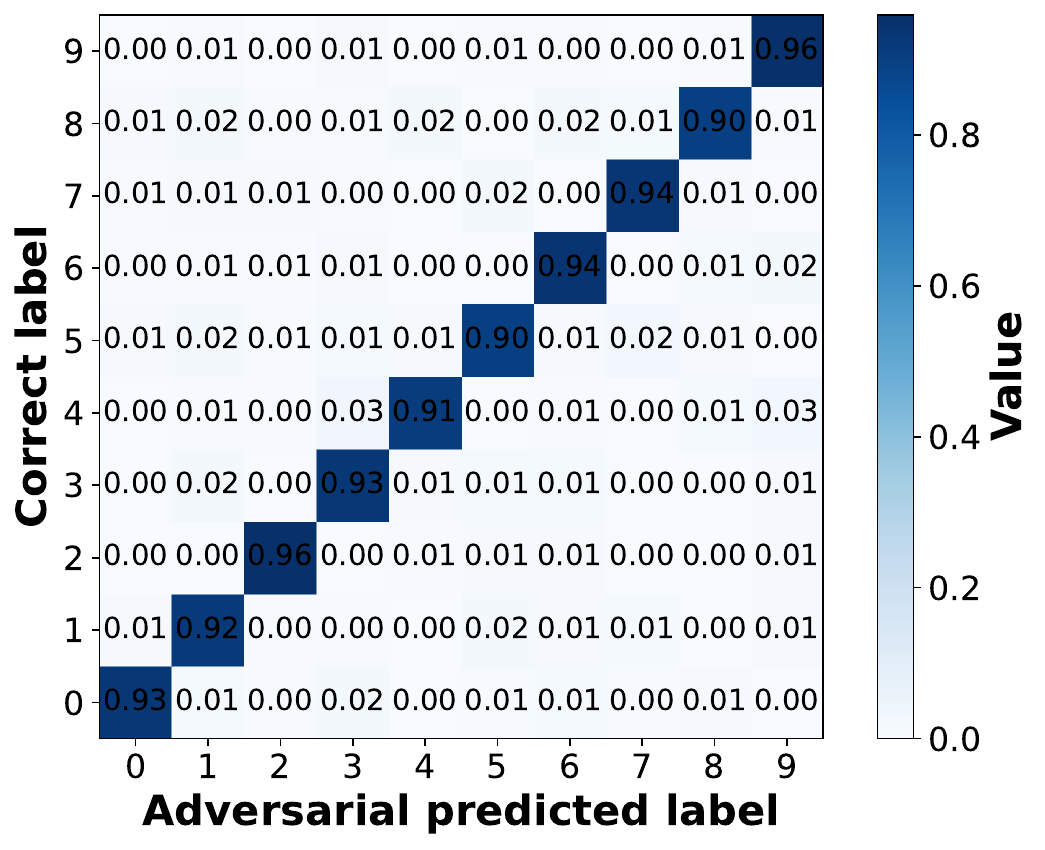}\label{fig:labelsb} 
	\caption{Adversarial predicted labels and correct labels for and 1000 poison samples on CIFAR-10. For poison samples, adversarial predictions predominantly cause the model to revert to the original correct labels, demonstrating a strong tendency to restore semantic consistency.}
	\label{fig:labels} 
\end{figure}

\section{Observations and Analysis}
In this section, we showcase key observations based on boundary adversarial attacks and provide our analysis for the subsequent method design.

\Tref{tab:boundary} presents the average decision boundary for 1000 clean samples and 1000 poison samples across seven backdoor attack types (BadNets \cite{gu2017badnets}, Blended \cite{chen2017targeted}, Dynamic \cite{nguyen2020input}, CL \cite{turner2019label}, LF \cite{zeng2021rethinking}, SSBA \cite{li2021invisible}, TrojanNN \cite{liu2018trojaning}) on CIFAR-10, CIFAR-100, and Tiny ImageNet-200. 
A key finding is that poisoned samples exhibit significantly larger distances from decision boundaries compared to clean samples. This is evident from the substantially larger perturbation magnitudes (measured via \(L_2\) norms) required to induce misclassification. These results align with previous research 
\cite{rajabi2023mdtd,su2024model}, which shows that backdoor attacks reshape decision boundaries, causing poison samples to cluster within an expanded high-confidence region.
For backdoored models, we observe a striking difference in the behavior of adversarially perturbed samples:
Clean samples exhibit uniformly distributed adversarial labels, as shown in \Fref{fig:labels} (left).
Poison samples, however, predominantly revert to their original semantic labels, as shown in \Fref{fig:labels} (right).
This phenomenon suggests that adversarial perturbations disrupt backdoor-specific features, leading poison samples to recover their original classification.
Notably, this observation holds regardless of the target label or the trigger mechanism.

\begin{figure*}[t]
	\centering
 	\includegraphics[width=.98\linewidth] {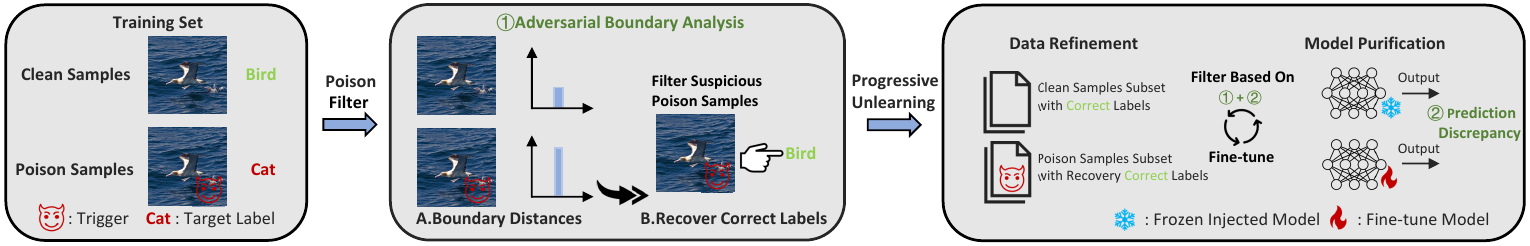}
	\caption{\textbf{Overview of \name framework.} The framework operates in two phases. \textbf{Phase 1}: Suspicious poisoned samples are identified via adversarial boundary analysis and re-labeled with their correct semantic labels. The model is then fine-tuned using these re-labeled samples, while retaining the frozen infected model for later comparison. \textbf{Phase 2}: By applying adversarial boundary analysis on the frozen infected model and leveraging dual-model prediction discrepancies, the framework progressively refines the dataset and purifies the model, effectively decoupling backdoor correlations while preserving clean-task performance.} 
    \label{fig:main} 
\end{figure*}

To further analyze this effect, we visualize the feature representations of: a clean sample $x$, its corresponding poisoned version $\hat{x}$, and the boundary adversarially perturbed version $\hat{x} + \delta$.
As shown in \Fref{fig:feature}: the cosine similarity between the clean sample $x$ and the poison sample $\hat{x}$ is relatively low (0.52), indicating a feature space shift due to the trigger. After applying adversarial perturbations, the cosine similarity between $\hat{x} + \delta$ and $x$ increases significantly to 0.88, demonstrating a realignment with the clean feature distribution.
Backdoor attacks inject trigger patterns, forming a shortcut mapping that forces poisoned samples to rely on trigger features rather than semantic features. 
This results in: 1) Poison samples deviating from the natural feature distribution of their original class. 2) The creation of a distinct ``backdoor subspace” where these samples reside.
To induce misclassification, adversarial perturbations must push samples out of their feature subspace. Since backdoored models learn high-confidence regions around the target label, larger perturbations are required to cross decision boundaries.
Moreover, adversarial perturbations may disrupt trigger features by modifying pixel patterns or textures, making them unrecognizable to the model. This collapses the backdoor effect, forcing poisoned samples to realign with clean features, which remain linked to their original class labels.


\section{Method}
\subsection{Overview}

Inspired by the observation based on adversarial boundary analysis, we propose \name, a unified framework that integrates backdoor false correlation decoupling and progressive purification. As shown in \Fref{fig:main}, this approach leverages adversarial boundary analysis alongside a dual-model prediction discrepancy feedback mechanism to guide progressive data refinement and model purification. By doing so, \name can effectively eliminate the false correlations introduced by backdoors while preserving the model’s performance on clean tasks. The method consists of the following key steps.

\subsection{Technical Details of The Design}
\noindent\textbf{Phase 1: Initial Potential Poison Samples Filtering.} 
Given a backdoor-infected model \( f(\cdot;\theta) \), we first identify candidate poison samples by analyzing their adversarial boundary distances. For each training sample \( x_i \in D_{\text{train}} \), we compute the mimimal perturbation \( \delta_i \) required to misclassify \( x_i \) via a boundary adversarial attack. The magnitude of this perturbation, measured by the \( l_2 \) norm, is: \(l(x_i) = \|\delta_i\|_2.\).
Larger values of \(l(x_i)\) indicate a higher likelihood that the sample is poisoned. 
Next, we sort all samples in \textit{descending order} based on \( l(x_i) \) and select the top \( k_p^0\% \) as the initial candidate poison set. The labels of these samples are then replaced with their adversarial predicted labels:
\[
\hat{y}_i = \arg\max f(x_i + \delta_i; \theta).
\]
This process forms the initial poison set \( D_{\text{poison}}^0\). To disrupt the false correlation between triggers and target labels, we fine-tune the model using the initial poison set: 
\[
\theta^t = \arg\min_{\theta^0} \mathbb{E}_{(x_i,\hat{y}_i) \in D^0_{\text{poison}}} \mathcal{L}(f(x_i;\theta^0), \hat{y}_i).
\]  
Simultaneously, the original infected model \( f(\cdot;\theta^0) \) is frozen for later comparison. 

After this initial fine-tuning, poisoned samples exhibit significantly larger prediction discrepancies between the frozen infected model and the progressively purified model, whereas clean samples show minimal changes. This forms the foundation of our dual-model prediction discrepancy feedback mechanism, which iteratively refines the training data and enhances model purification.


\noindent\textbf{Phase 2: Progressive Data Refinement and Model Purification.}
To further distinguish poisoned samples from clean ones, we introduce two complementary metrics.
For each training sample \( x_i \in D_{\text{train}} \), we first compute the normalized adversarial boundary distance (BD), \ie, 
\[
{BD}(x_i) = \frac{l(x_i)}{l_{\text{max}}},
\] 
where \( l(x_i) \) represents the adversarial perturbation magnitude required to misclassify \( x_i \) and \( l_{\text{max}} \) denotes the maximum perturbation across all samples. This measures how far a sample is from the decision boundary, with poison samples generally requiring larger perturbations.
The second metric is the prediction divergence (PD), \ie, 
\[
{PD}(x_i) = 1 - \cos\left(f(x_i;\theta_0), f(x_i;\theta^t)\right),
\] 
where \( \theta_0 \) is the original infected model and \( \theta^t \) is the model at iteration \( t \). 
This captures the activation shifts caused by purification, with poison samples showing higher divergence due to the removal of backdoor triggers.
We define a composite score to dynamically balance these metrics:
\[ 
S(x_i) = (1-\omega^t) \cdot \text{BD}(x_i) + \omega^t \cdot \text{PD}(x_i), 
\] 
where the dynamic weight \( \omega^t \) increases over iterations \( t \), shifting emphasis from boundary distance (BD) in early stages to prediction divergence (PD) in later stages.

To balance precision and coverage during purification, we employ an adaptive sample selection strategy. Training samples are ranked by their composite score \( S(x_i) \). Besides, the top \( k_p^t\% \) are assigned to the poison set \( D_{\text{poison}}^t \) and replace their labels with the adversarial predicted labels $\hat{y}_i = \arg\max f(x_i + \delta_i; \theta)$, while the bottom \( k_c^t\% \) are assigned to the clean set \( D_{\text{clean}}^t \). 
The model is then iteratively fine-tuned on the union of the clean and poisoned sets:
\[
\theta^{t+1} = \arg\min_{\theta^t} \mathbb{E}_{(x_i,y_i) \in D_{\text{poison}}^t \cup D_{\text{clean}}^t} \mathcal{L}(f(x_i;\theta^t), y_i).
\]

\noindent\textbf{Hyper-parameters Configuration.} 
To ensure a balanced trade-off between precision and coverage, we dynamically adjust the proportion of selected poisoned and clean samples over iterations:
\[
k_p^t = k_p^0 + \frac{t}{T}\left(k_p^T - k_p^0\right),
\\
k_c^t = k_c^0 + \frac{t}{T}\left(k_c^T - k_c^0\right).
\]
Specifically, \(k_p^0\) is set to 0.5\%, \(k_p^T\) is set to 1\%, \(k_c^0\) is set to 2.5\%, and \(k_c^T\) is set to 5.0\%. In practice, an attacker requires a certain poisoning rate to ensure the attack's effectiveness, and thus we assuming a 1\% screened subset of poison samples. And a larger proportion of clean samples is beneficial for restoring the model's performance on the original task.
In addition, we adopt dynamic weight scheduling on \(\omega^t\), \ie,  
\[
\omega^t = \omega^0 + \frac{t}{T}(\omega^T - \omega^0),
\]  
where \(t\) denotes the iteration index, $\omega^0=0.3$ and $\omega^T=0.6$. 
In early iterations, filtering relies more on boundary distances (BD) to detect poisoned samples. In later iterations, prediction divergence (PD) becomes the dominant criterion as purification progresses.
    

\Aref{alg:ABU} shows the overview of \name. During each iteration, the model undergoes fine-tuning for 5 epochs using the Adam optimizer (\(\beta_1=0.9, \beta_2=0.999\)) with a fixed learning rate of \(lr=5 \times 10^{-5}\). And, we fine-tune $T=20$ iterations in total.

\begin{algorithm}[!h]
    \caption{Backdoor Unlearning Based on Adversarial Boundary Analysis (\name).}
    \label{alg:ABU}
    \renewcommand{\algorithmicrequire}{\textbf{Input:}}
    \renewcommand{\algorithmicensure}{\textbf{Output:}}
    
    \begin{algorithmic}[1]
        \REQUIRE 
        Infected model $f( ; \theta)$, 
        Training data $D_{\text{train}} = \{(x_i, y_i)\}_{i=1}^n$,
        Max Iterations $T$,
        Proportion of dataset \( k_p^0/k_p^T \) and \( k_c^0/k_c^T \),
        Dynamic Weight \(\omega^t\)
        
        \ENSURE Purified model $f( ; \theta^T)$
        
        \STATE \textbf{Phase 1: Initial Poison Filter}
        \FOR{each $(x_i, y_i) \in D_{\text{train}}$}

            \STATE $l(x_i) \gets \|\delta_i\|_2, l_{\max} \gets \max(l_{\max},l(x_i))$
        \ENDFOR
        \STATE Sort $D_{\text{train}}$ by $l(x_i)$ descending
        \STATE $D_{\text{poison}}^0 \gets$ top $k_p^0\%$ samples$, \hat{y}_i = \arg\max f(x_i + \delta_i; \theta_0)$
        \STATE Frozen $\theta^0$ as \( f(\cdot;\theta^0) \)
        \STATE $\theta^1 \gets \text{Fine-tune}(\theta_0, D_{\text{poison}}^0)$
        
        \STATE \textbf{Phase 2: Progressive Backdoor Unlearning}
        \FOR{$t = 1, ..., T$}
                \FOR{each $x_i \in D_{\text{train}}$}
                \STATE $BD(x_i) \gets \frac{l(x_i)}{l_{\text{max}}}$
                \STATE $PD(x_i) \gets 1 - \cos(f(x_i;\theta^0), f(x_i;\theta^t))$ 
                \STATE $S(x_i) \gets (1-\omega^t)BD(x_i) + \omega^t PD(x_i)$
            \ENDFOR
        
            
            \STATE Sort $D_{\text{train}}$ by $S(x_i)$ descending
            \STATE Update poison set: $D_{\text{poison}}^t \gets$ top $k_p^t\%$ samples$, \hat{y}_i = \arg\max f(x_i + \delta_i; \theta_0)$
            \STATE Update clean set: $D_{\text{clean}}^t \gets$ bottom $k_c^t\%$ samples
            \STATE $\theta^{t+1} \gets \text{Fine-tune}(\theta^t, D_{\text{poison}}^t \cup D_{\text{clean}}^t)$
        \ENDFOR 
    \end{algorithmic}
\end{algorithm}

\begin{figure*}[t]
  \centering
  \begin{subfloat}{
    \includegraphics[width=0.75\textwidth]{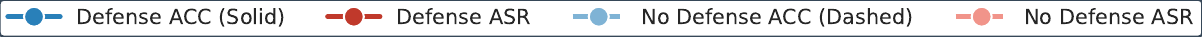}}
  \end{subfloat}
  \\
  \begin{subfloat}{
    \includegraphics[width=0.22\textwidth]{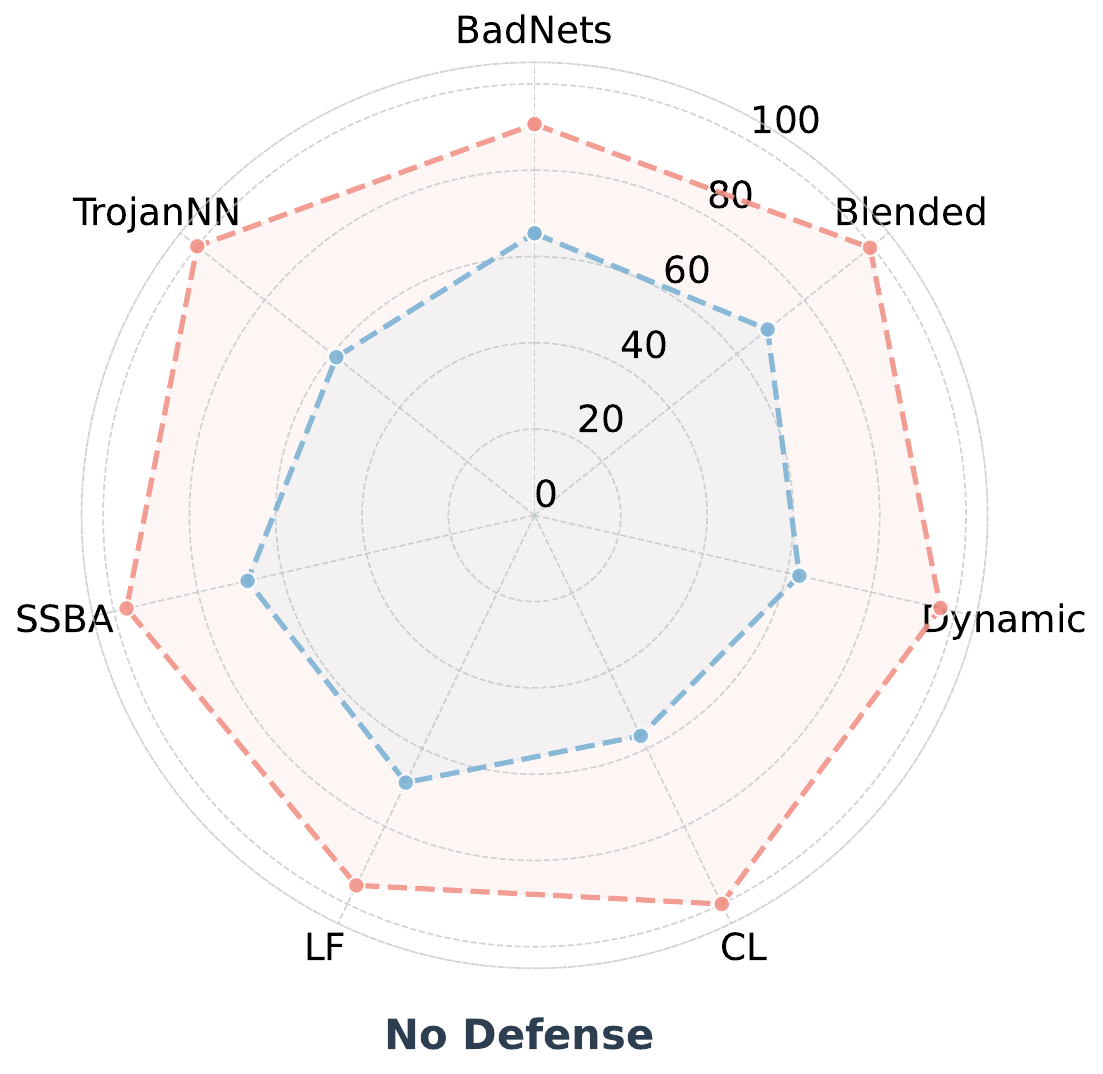}}    
  \end{subfloat}
  \begin{subfloat}{
    \includegraphics[width=0.22\textwidth]{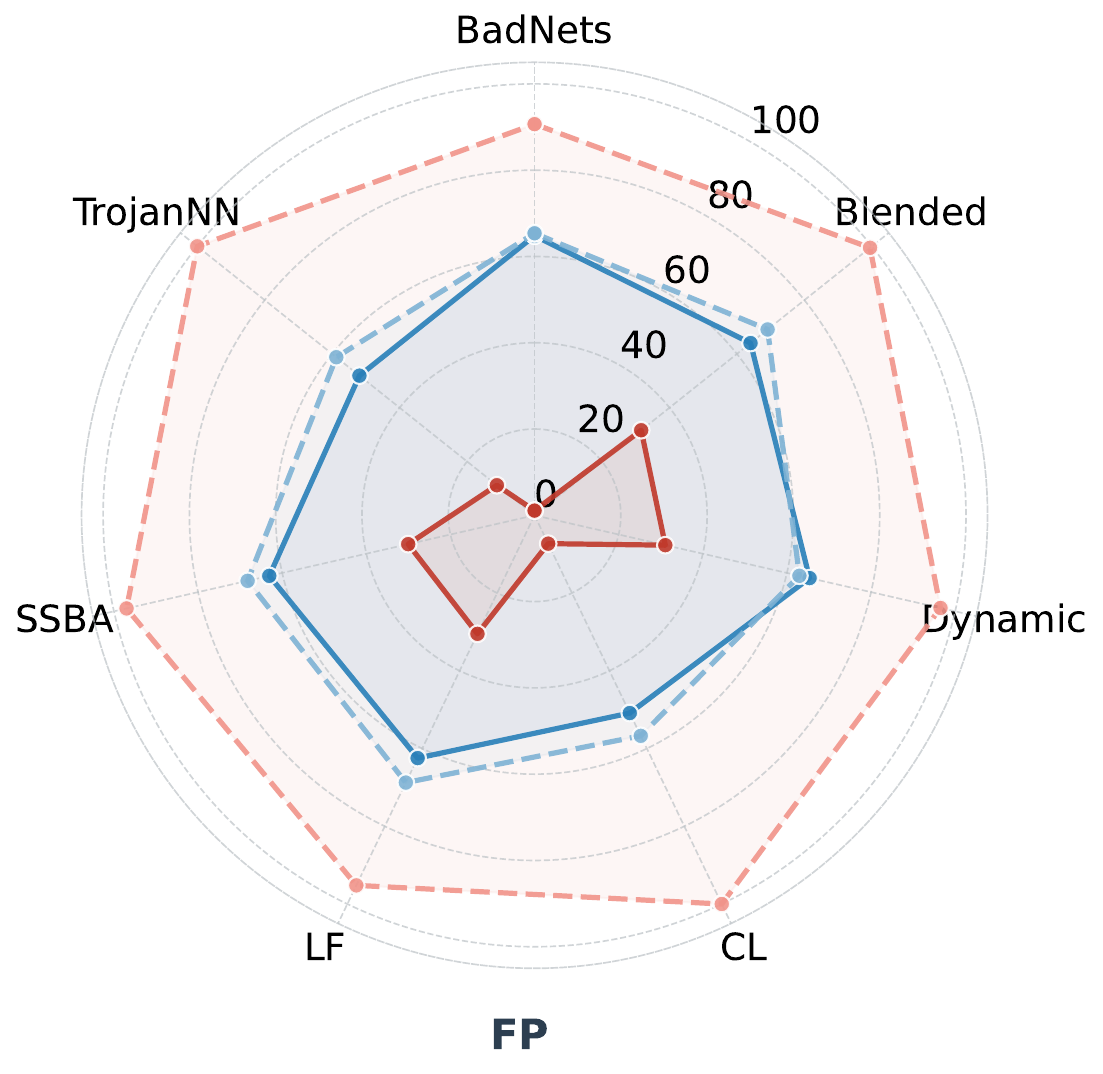}}
  \end{subfloat}
  \begin{subfloat}{
    \includegraphics[width=0.22\textwidth]{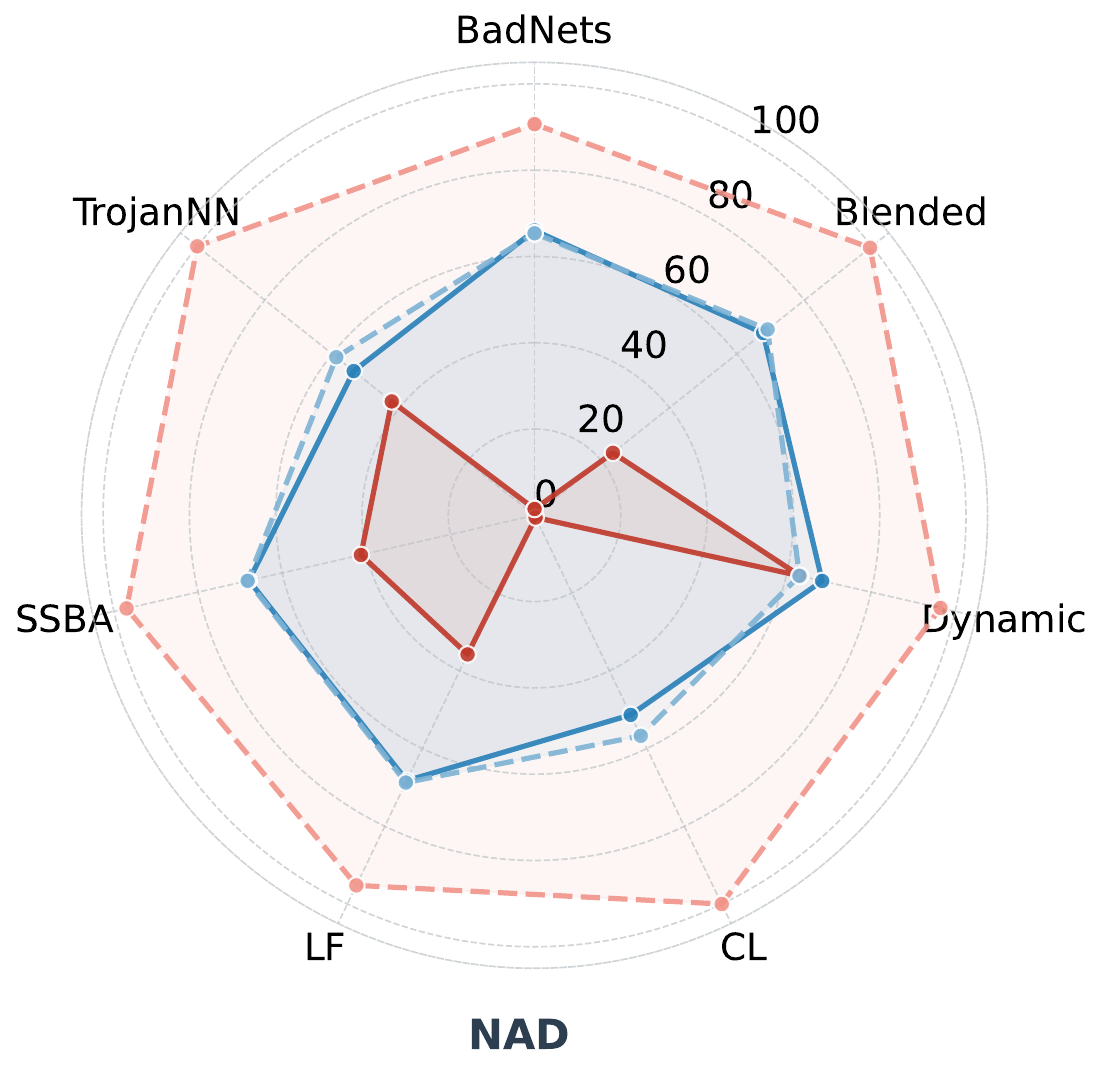}}    
  \end{subfloat}
  \begin{subfloat}{
    \includegraphics[width=0.22\textwidth]{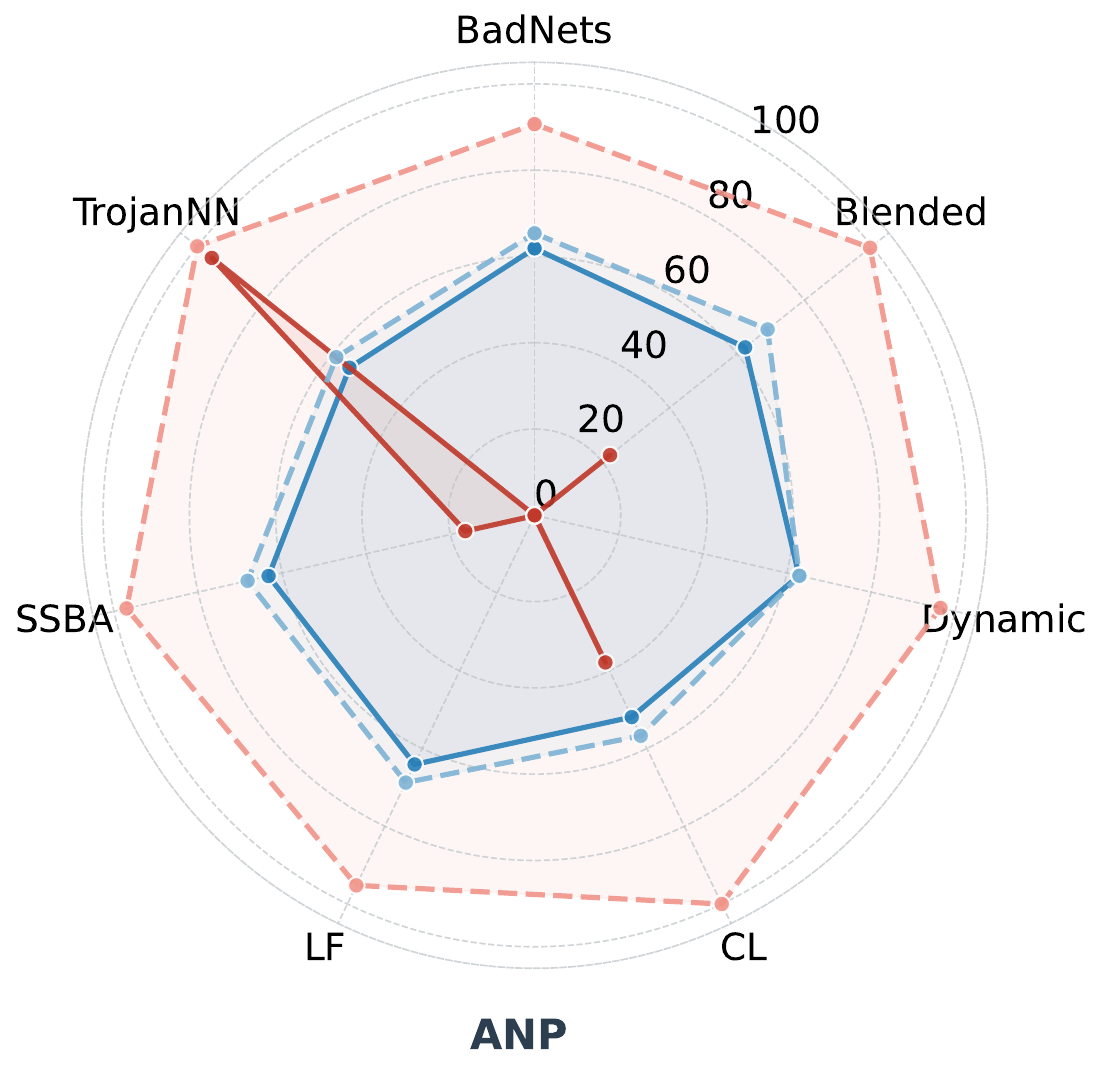}}    
  \end{subfloat}

  \begin{subfloat}{
    \includegraphics[width=0.22\textwidth]{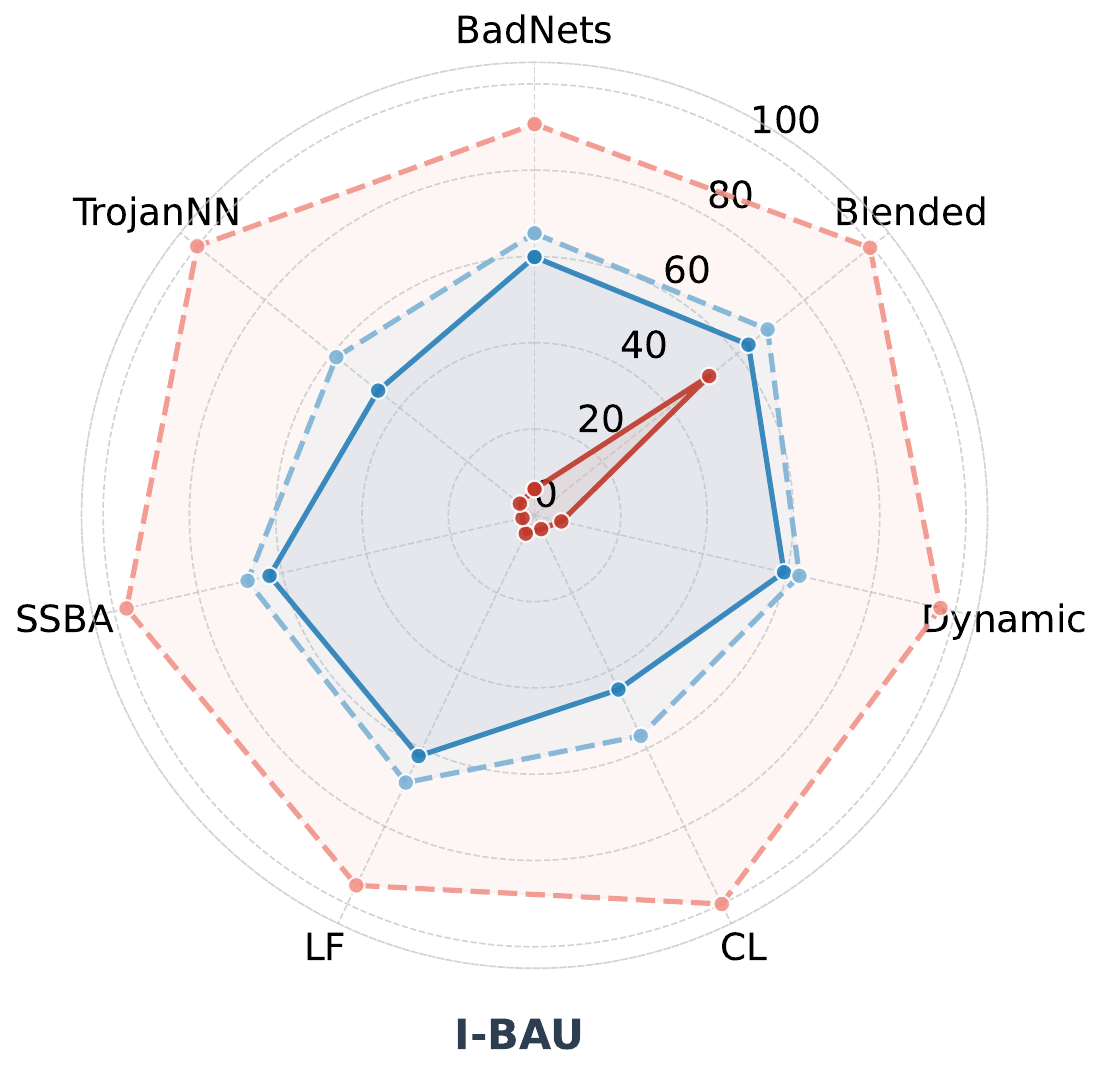}}
  \end{subfloat}
  \begin{subfloat}{
    \includegraphics[width=0.22\textwidth]{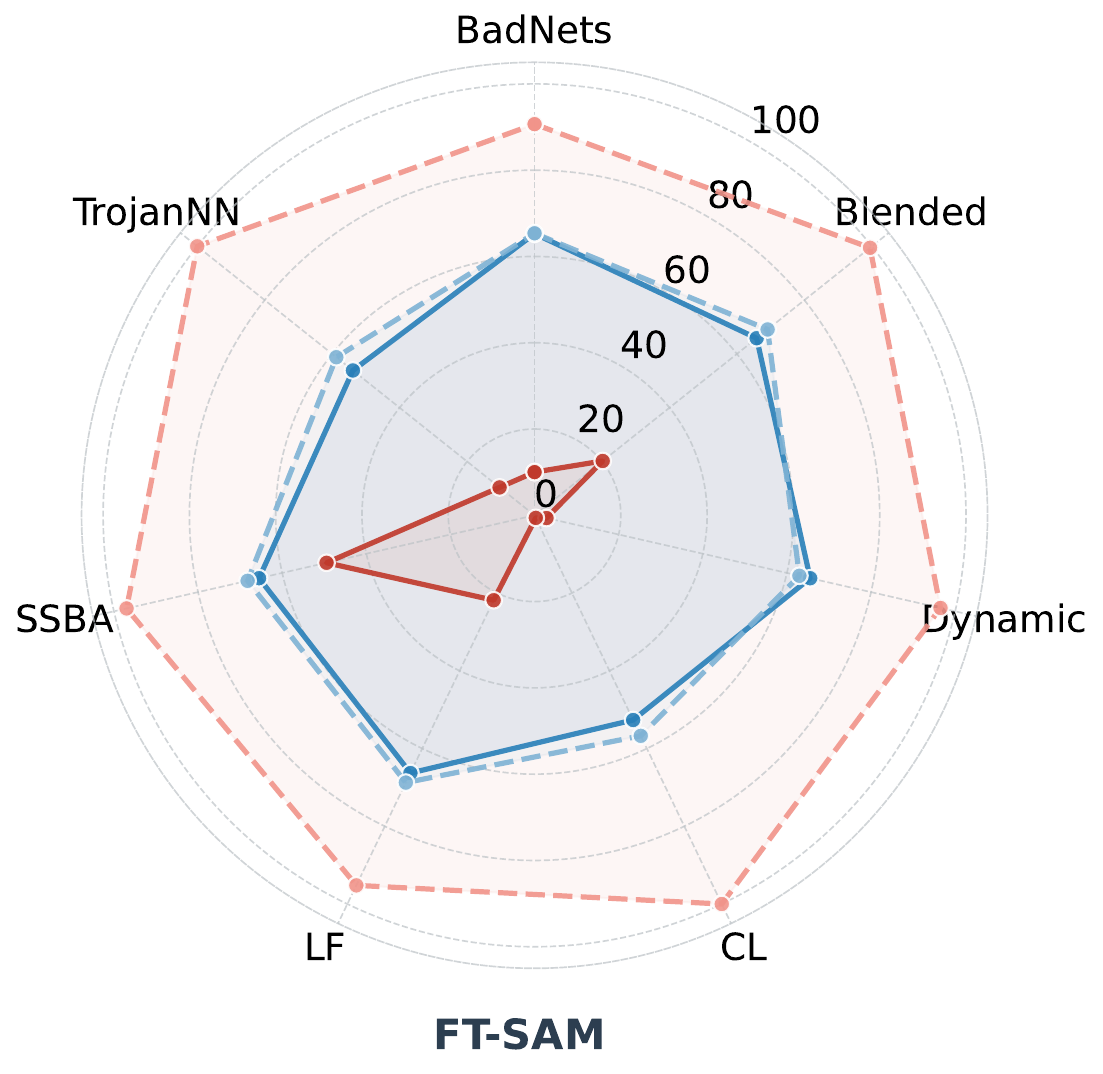}}    
  \end{subfloat}
  \begin{subfloat}{
    \includegraphics[width=0.22\textwidth]{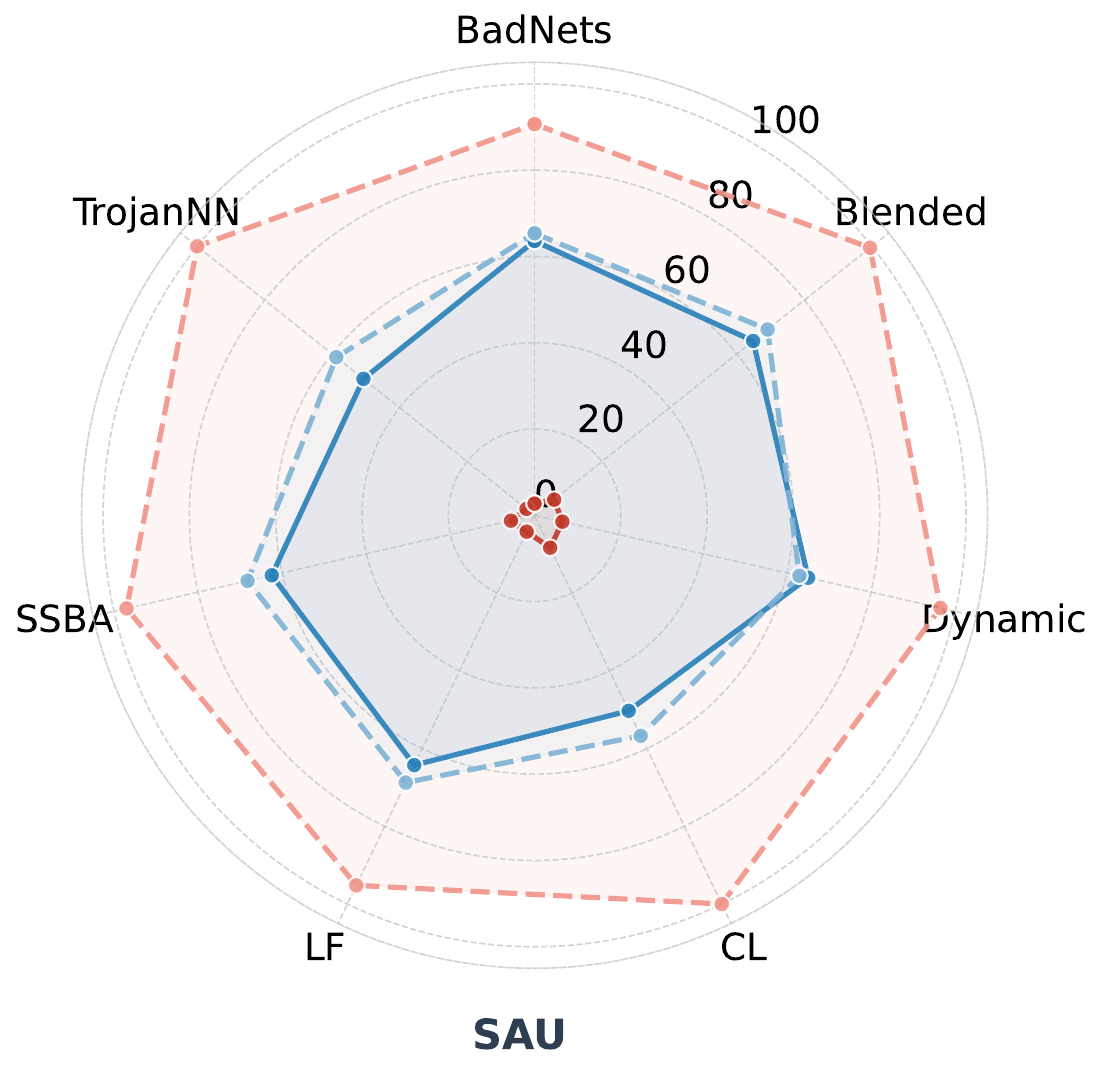}}    
  \end{subfloat}
  \begin{subfloat}{
    \includegraphics[width=0.22\textwidth]{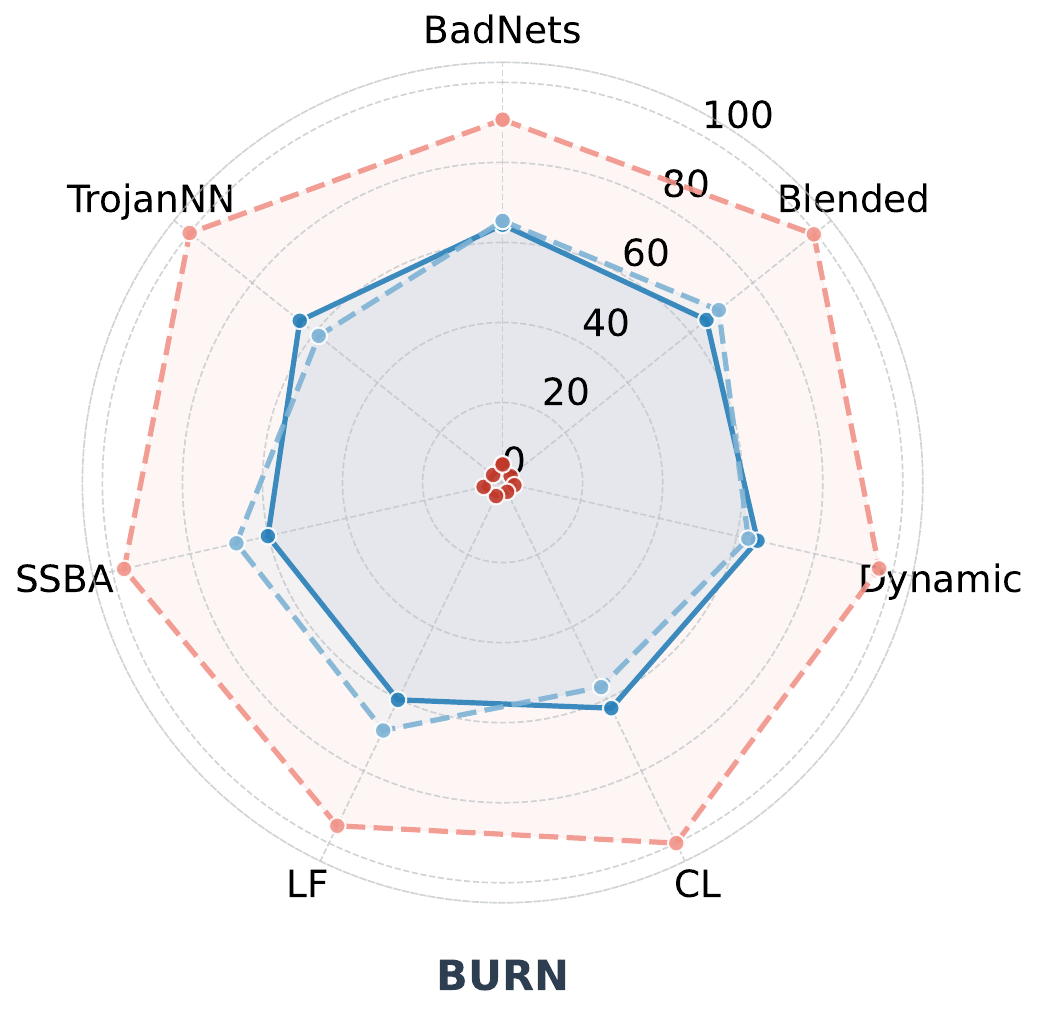}}    
  \end{subfloat}
  
 	\caption{The radar plots visualize ACC and ASR trends for each method, where solid lines represent defended models and dashed lines indicate performance without defense. Lower ASR and higher ACC indicate stronger defense efficacy.}
	\label{fig:radar} 
\end{figure*}

\begin{figure*}[t]
  \centering

  \begin{subfloat}{
    \includegraphics[width=0.13\textwidth]{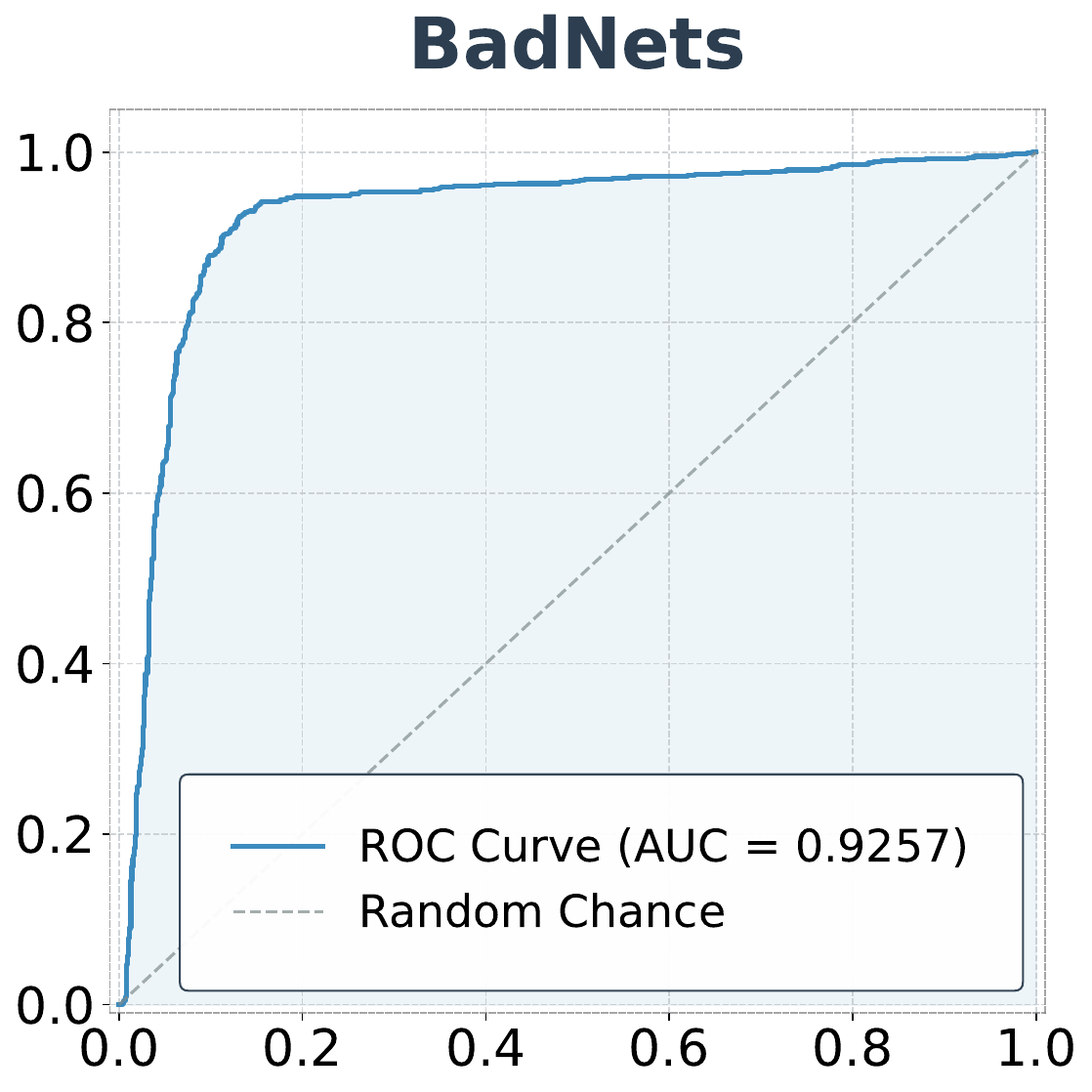}}
  \end{subfloat}
  \begin{subfloat}{
    \includegraphics[width=0.13\textwidth]{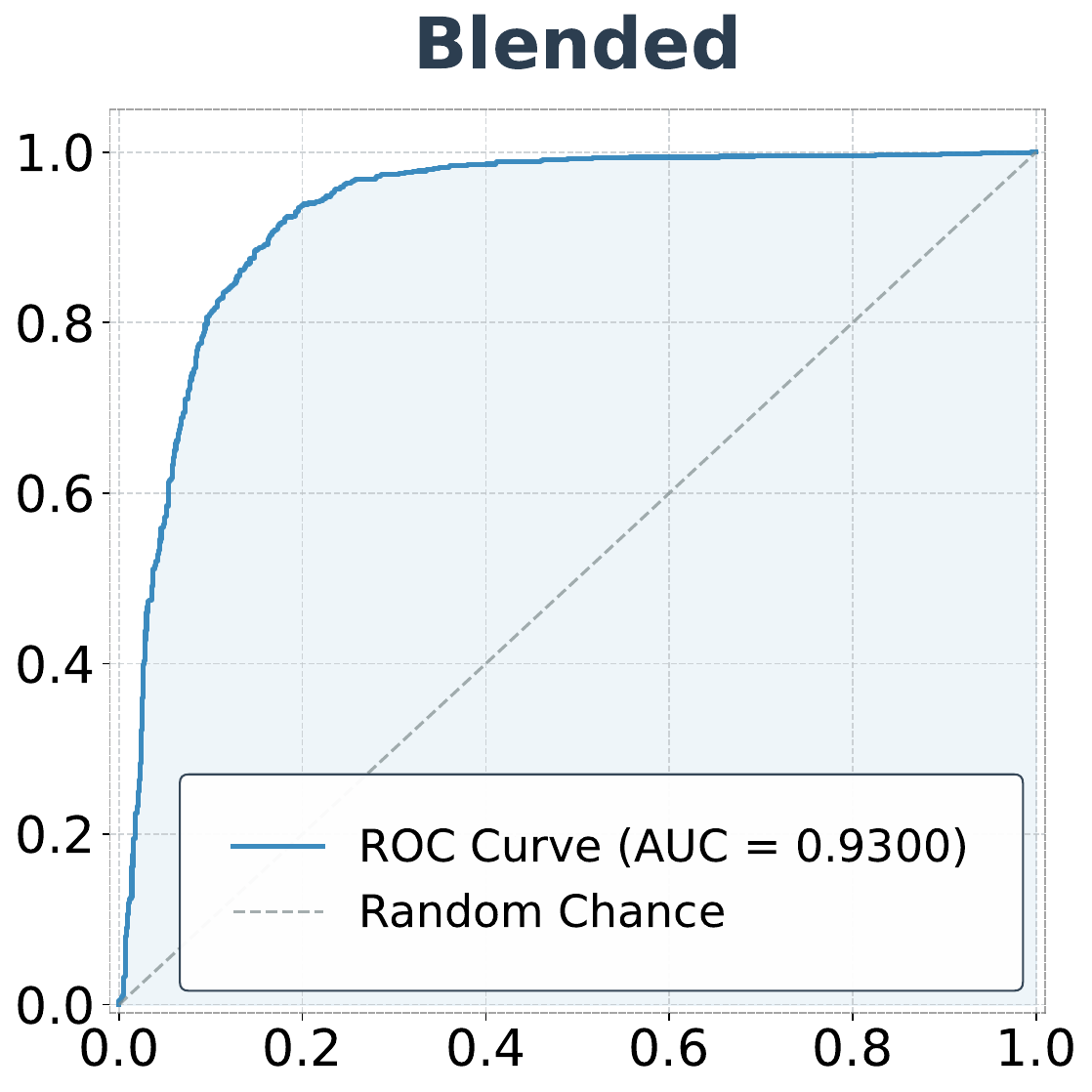}}    
  \end{subfloat}
  \begin{subfloat}{
    \includegraphics[width=0.13\textwidth]{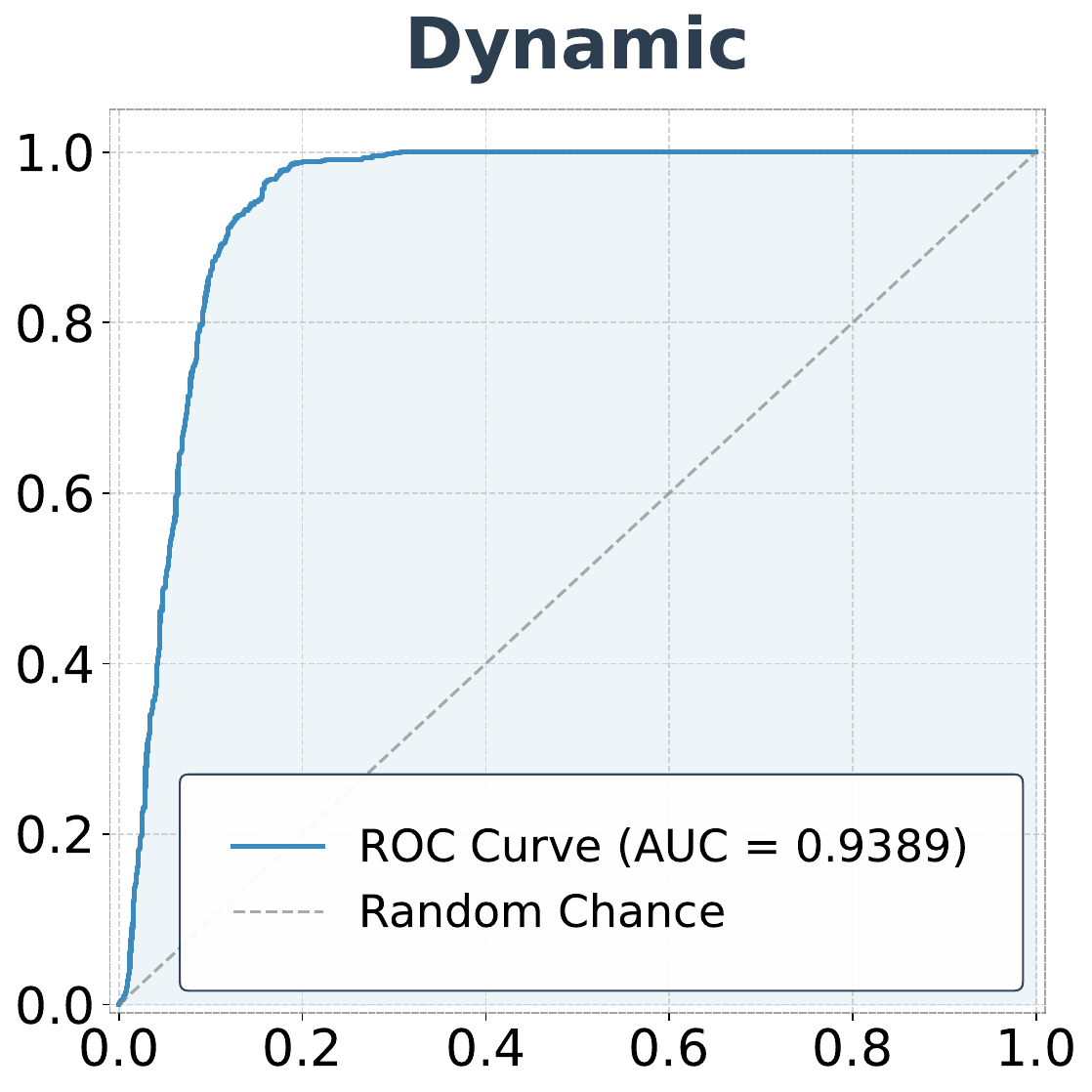}}    
  \end{subfloat}
  \begin{subfloat}{
    \includegraphics[width=0.13\textwidth]{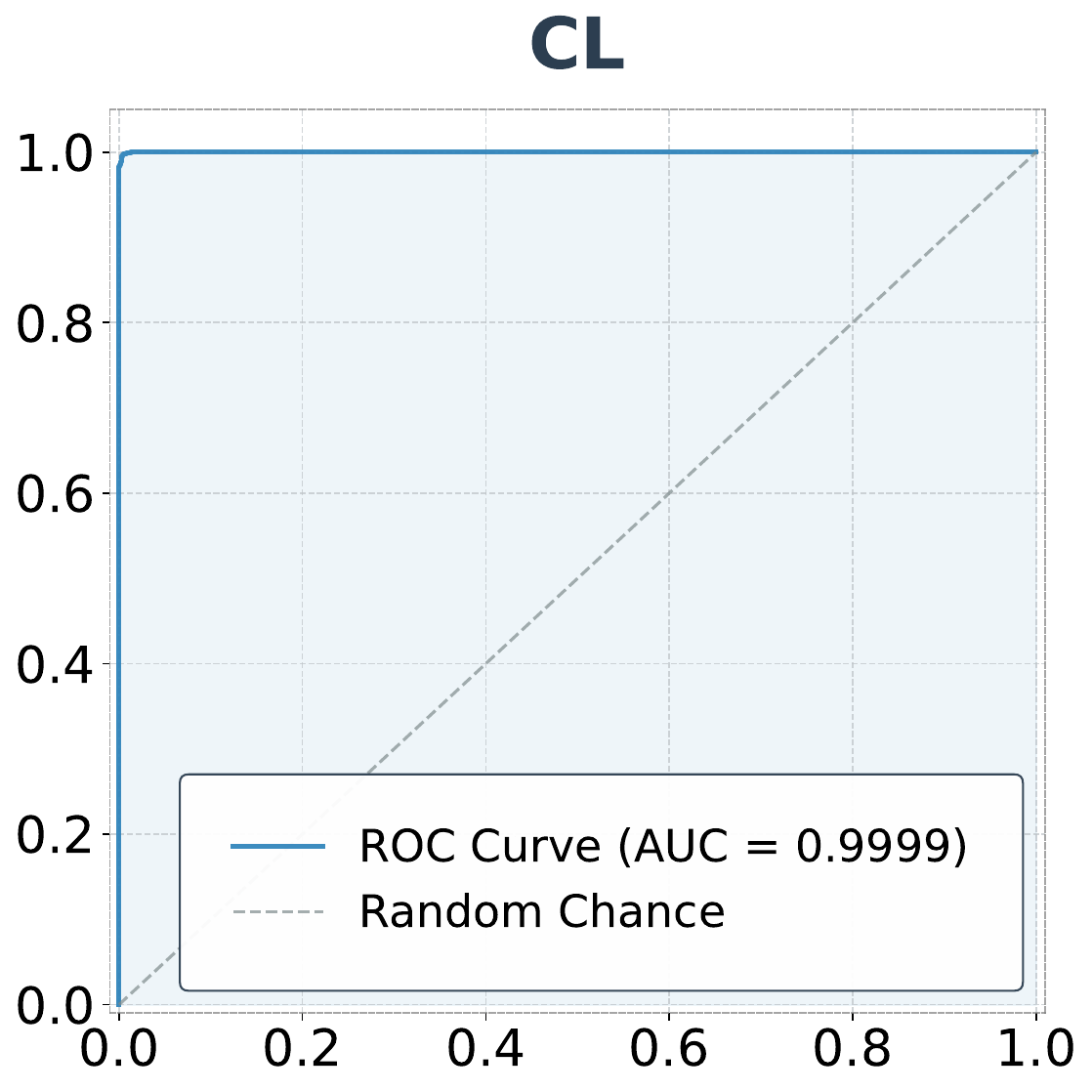}}    
  \end{subfloat}
  \begin{subfloat}{
    \includegraphics[width=0.13\textwidth]{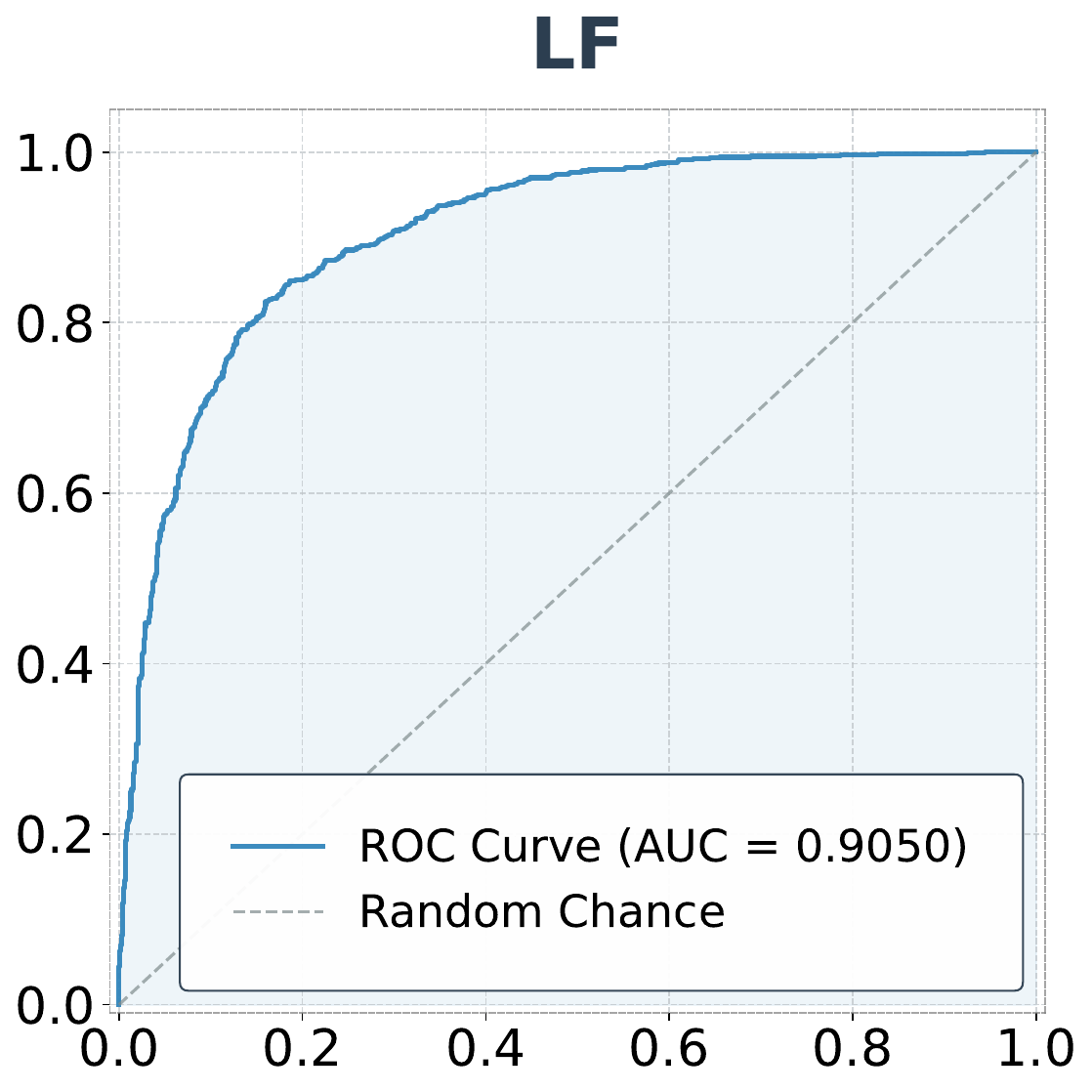}}
  \end{subfloat}
  \begin{subfloat}{
    \includegraphics[width=0.13\textwidth]{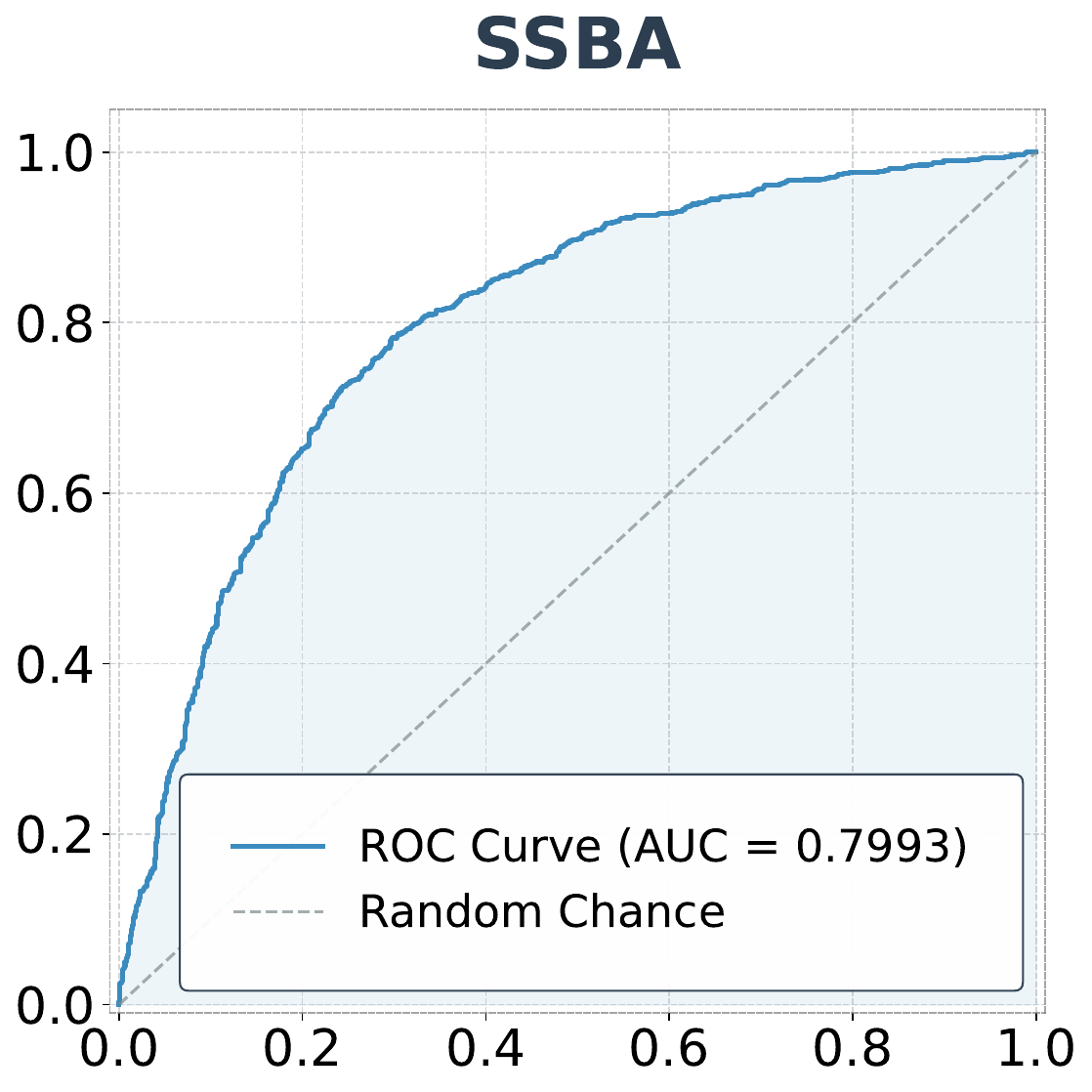}}    
  \end{subfloat}
  \begin{subfloat}{
    \includegraphics[width=0.13\textwidth]{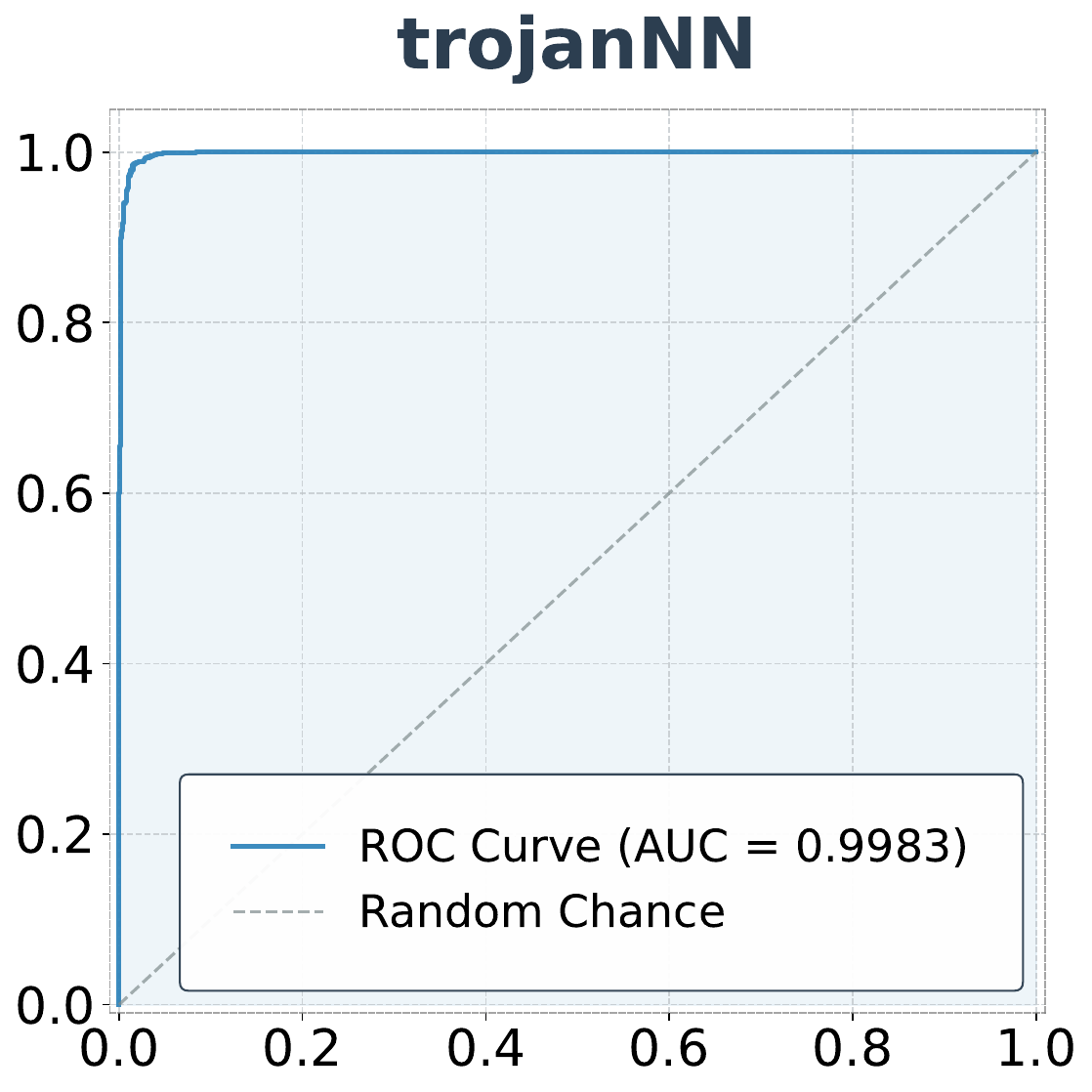}}    
  \end{subfloat}
 \vspace{-.5em}
 	\caption{
     Performance of poison sample filtering using adversarial boundary analysis. The receiver operating characteristic (ROC) curves illustrate the trade-off between the false positive rate (X-axis, clean samples misclassified as poison) and the true positive rate (Y-axis, correctly identified poison samples) across various backdoor attacks. The area under the ROC curve (AUROC) quantifies the effectiveness of distinguishing poison from clean samples, with higher values indicating better detection performance.
     }
   \vspace{-1em}
	\label{fig:auroc} 
\end{figure*}

\section{Experiment}
\subsection{Experimental Setting}
\noindent\textbf{Datasets and Architectures.} 
The datasets include CIFAR-10 \cite{krizhevsky2009learning}, CIFAR-100 \cite{krizhevsky2009learning}, and Tiny ImageNet-200 \cite{deng2009imagenet}. To assess the generalizability of \name across different network architectures, we employ Pre-Activation ResNet-18 \cite{he2016identity} as the architecture for CIFAR-10, and Pre-Activation ResNet-34 \cite{he2016identity} for CIFAR-100 and Tiny ImageNet-200.

\noindent\textbf{Attack Baseline.} For the attack baselines, we evaluate our method against seven backdoor attacks, including BadNets \cite{gu2017badnets}, Blended \cite{chen2017targeted}, Dynamic \cite{nguyen2020input}, CL \cite{turner2019label}, LF \cite{zeng2021rethinking}, SSBA \cite{li2021invisible}, TrojanNN \cite{liu2018trojaning}.
We follow an open-sourced benchmark BackdoorBench \cite{wu2022backdoorbench} for the training settings of these attacks and conduct all-to-one attacks by default. 
To ensure the attacks’ strength, the poisoning ratio adopts 10\% by default. 

\noindent\textbf{Defense Baseline.} We compare our \name approach with 7 state-of-art backdoor defense methods: 1) Fine-pruning (FP) \cite{liu2018fine}, 2) Neural Attention Distillation (NAD) \cite{li2021neural}, 3) Adversarial Neuron Pruning (ANP) \cite{wu2021adversarial}, 4) Implicit Backdoor Adversarial Unlearning (I-BAU) \cite{zeng2021adversarial}, 5) FT-SAM \cite{zhu2023enhancing}, 6) Shared Adversarial Unlearning (SAU) \cite{wei2023shared}. 
To ensure reproducibility and maintain a fair comparison, we employ the official default implementations provided in widely recognized open-source benchmarks \cite{wu2022backdoorbench}.
Regarding the clean extra data, we follow the same protocol of these methods: the extra clean data is randomly selected from clean training data, taking about 5\% of all training data.

\noindent\textbf{Evaluation Metrics.} 
We assess defense method efficacy through two critical metrics: the attack success rate (ASR), quantifying the percentage of adversarial samples incorrectly assigned to the attacker-specified target label, and the accuracy on clean samples (ACC), reflecting the model's retained classification performance on unperturbed inputs.

\subsection{Effectiveness Evaluation} 
We conduct a comprehensive evaluation of \name's performance against seven backdoor attacks across three datasets. Due to space constraints, we focus our analysis on the CIFAR-100 results in \Fref{fig:radar}, with comprehensive results for all datasets provided in the supplementary material.  \Fref{fig:radar} illustrates the ACC-ASR trade-offs across all methods, reveal that \name consistently suppresses attack success rates to low levels while maintaining model utility.

In summary, \name consistently demonstrates effectiveness against various attacks without relying on assumptions about specific trigger patterns, even in the absence of a clean auxiliary dataset.

\subsection{Ablation Study} 

\noindent\textbf{The Influence of the Hyper-parameters.}
%
Parameters \( k_p^0/k_p^T \) and \( k_c^0/k_c^T \) govern poisoned and clean subset proportions, respectively, crucially influencing method performance. We evaluate two configurations:  
1) Fixed \( k_c^{0,T} = \{2.5\%,5\%\} \) with \( k_p^{0,T} \in \{0.25\sim0.5\%,0.5\sim1\%,1\sim2\%\} \);  
2) Fixed \( k_p^{0,T} = \{0.5\%,1\%\} \) with \( k_c^{0,T} \in \{1.25\sim2.5\%,2.5\sim5\%,5\sim10\%\} \) against BadNets attack on CIFAR-100.  
As shown in Table \ref{tab:k}, elevated \( k_p \) reduces both ASR and accuracy, while increased \( k_c \) suppresses ASR while improving ACC. The \( k_p^T = 1\% \) threshold provides robust defense against adaptive poisoning rate manipulation, with \( k_c^T = 5\% \) ensuring conservative yet equitable benchmarking given potential incomplete defender datasets.

\begin{table}[t]
  \centering
  \footnotesize
  \setlength\tabcolsep{2pt}
  \caption{The influence of \( k_p^0/k_p^T \) and \( k_c^0/k_c^T \).}
  \vspace{-1em}
\begin{tabularx}{0.35\textwidth}{
  >{\bfseries}l
  *{10}{>{\centering\arraybackslash}X}
}
\hline
$k_p$ & $0.25\% \sim0.5\%$ & $0.5\% \sim1\%$ & $1\% \sim2\%$ \bigstrut\\
\hline
ACC & $65.55$  & $64.42$  & $62.78$  \bigstrut[t]\\
ASR & $5.78$  & $2.52$  & $2.28$  \bigstrut[b]\\
\hline

\end{tabularx}

\vspace{1em}

\begin{tabularx}{0.35\textwidth}{
  >{\bfseries}l
  *{10}{>{\centering\arraybackslash}X}
}
\hline
$k_c$ & $1.25\% \sim2.5\%$ & $2.5\% \sim5\%$ & $5\% \sim10\%$ \bigstrut\\
\hline
ACC & $61.86$  & $64.42$  & $66.55$  \bigstrut[t]\\
ASR & $5.78$  & $2.52$  & $1.80$  \bigstrut[b]\\
\hline
\end{tabularx}

  \label{tab:k}%
  \vspace{-1em}
\end{table}%

\subsection{More Anlysis} 
\noindent\textbf{The Effectivness of Filtering Potential Backdoor Samples.}
Since identifying potential backdoor samples is a crucial component of our defense framework, we conduct a comprehensive evaluation of filtering effectiveness across seven representative backdoor attacks: BadNets, Blended, Dynamic, CL, LF, SSBA, and TrojanNN, implemented on the CIFAR-100 dataset.
As illustrated in \Fref{fig:auroc}, we systematically analyze the adversarial boundary distance metric for distinguishing poisoned samples from clean samples. Our experimental results confirm that this metric achieves robust separability, effectively discriminating between clean and poisoned samples with high accuracy.

\begin{figure}[t]
	\centering
 	\includegraphics[width=.48\linewidth] {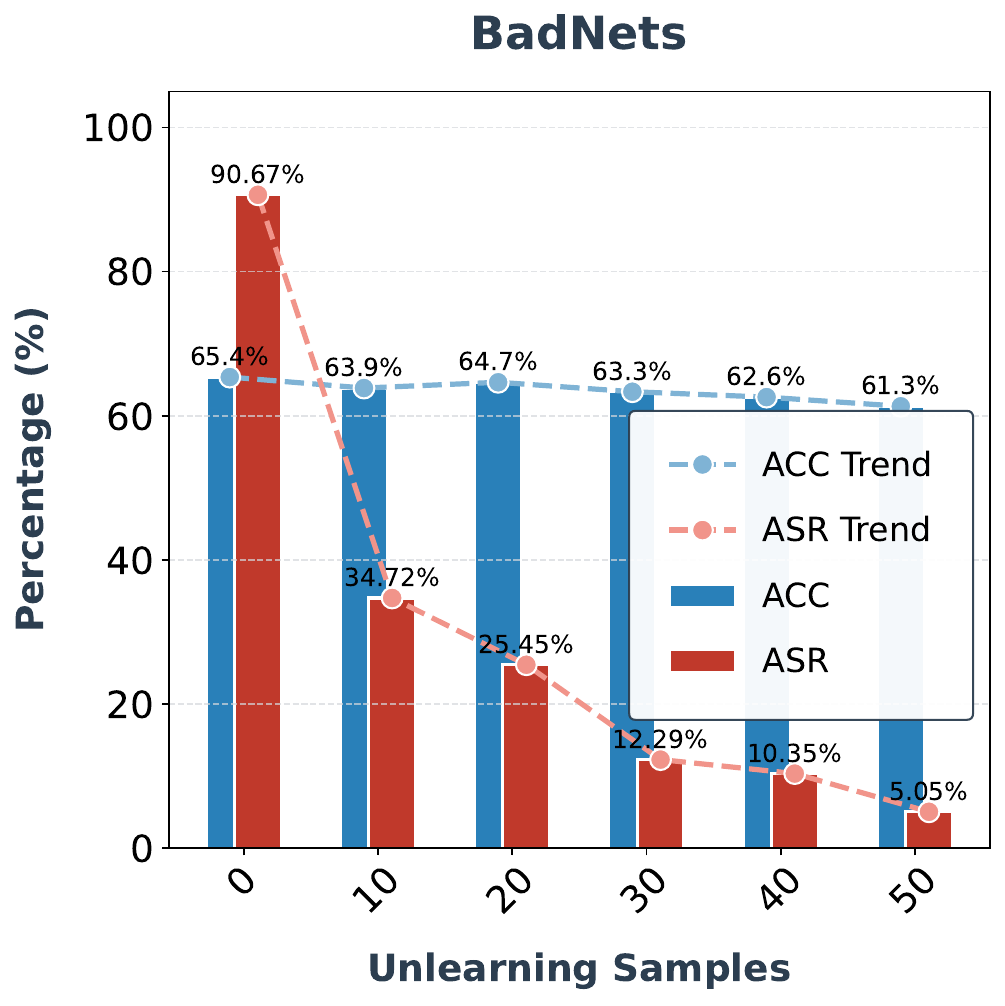}
        \includegraphics[width=.48\linewidth] {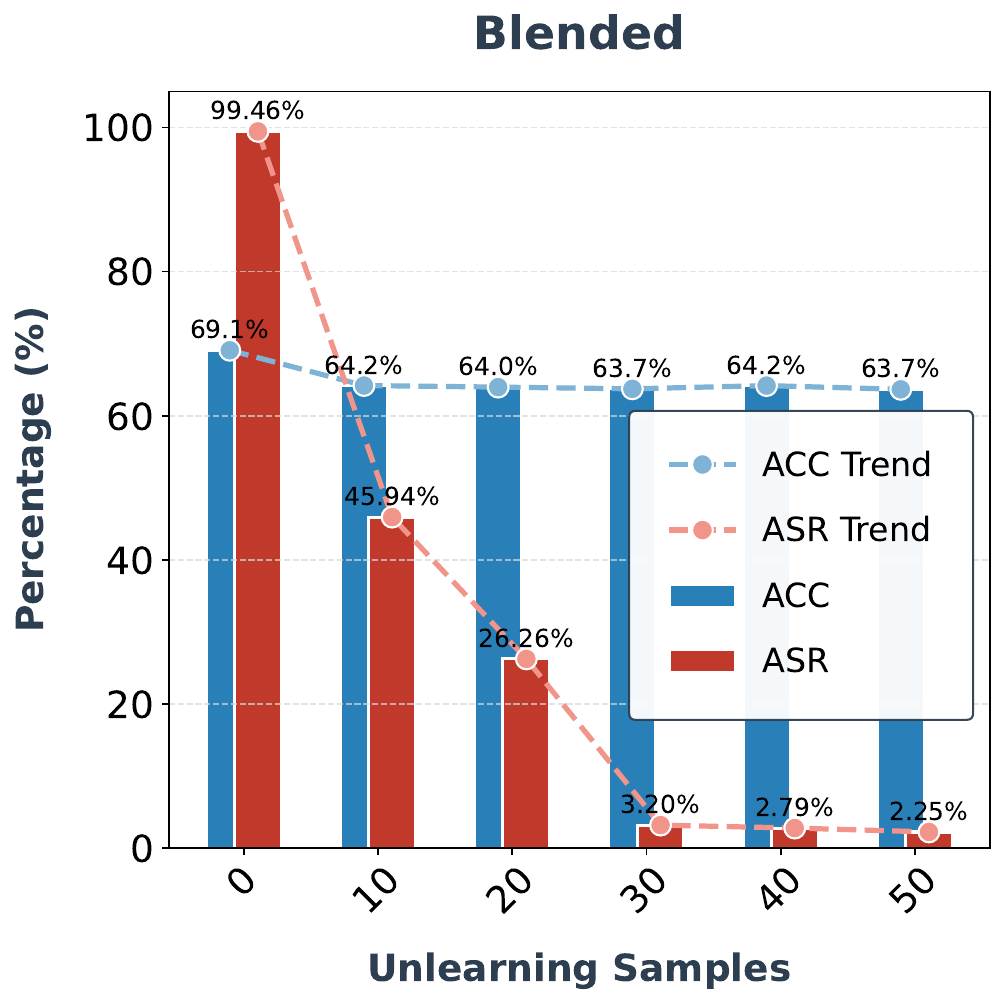}
 \vspace{-.5em}
	\caption{The trends of model accuracy (ACC) and attack success rate (ASR) as the number of unlearning poison samples increases for BadNets attack (left) and Blended attack(right).}
	\label{fig:samples} 
\end{figure}

\noindent\textbf{The Efficacy of Backdoor Mitigation via Unlearning.} To rigorously evaluate the capability of our unlearning framework in decoupling false trigger-target correlations, we developed a quantitative experiment to characterize the relationship between the number of identified poisoned samples and the model’s behavior on the CIFAR-100 dataset, measured by ACC and ASR. Specifically, we fine-tuned the model for one iteration using these samples with restored correct labels.
As demonstrated in \Fref{fig:samples}, our analysis reveals that effective backdoor suppression can be achieved without exhaustive detection of poisoned training instances. The unlearning protocol achieves dual objectives: (1) preserving model fidelity while (2) inducing exponential decay of attack potency, requiring identification of only limited poisoned samples ($\approx$ 30) for complete backdoor neutralization.

\section{Conclusions, Limitations, and Future Work}

In this paper, we present intriguing observations that poison samples exhibit greater distances from decision boundaries and that boundary adversarial attacks can counteract these trigger patterns to restore the correct labels. Building on these insights, we propose \name, a novel defense framework that integrates backdoor connection decoupling and progressive purification without relying on assumptions about specific trigger patterns, even without the need for a clean auxiliary dataset. 
However, \name’s effectiveness depends on the accurate detection of poisoned samples, which presents a potential limitation. To further enhance its robustness, integrating \name with state-of-the-art backdoor detection methods will be a key focus of our future work.

\small
\bibliographystyle{ieeenat_fullname}
\bibliography{main}

\end{document}